\documentclass[11pt,authoryear]{elsarticle}

\usepackage{microtype}

\usepackage{setspace}

\usepackage{amsthm}
\usepackage{amssymb} 

\usepackage{float} 

\usepackage{threeparttable}

\usepackage{graphicx}

\usepackage{subcaption} 

\usepackage{mathtools}

\usepackage{comment}

\usepackage[table]{xcolor}  
 
\usepackage{bm}
\usepackage{latexsym}
\usepackage{mathtools}

\usepackage{tikz}
\usepackage{tikz-cd}
\usetikzlibrary{shapes.geometric, arrows.meta, positioning, fit, shadows.blur, calc}
\tikzcdset{diagrams={nodes={inner sep=7.5pt}}}

\usepackage{pgfplots}
\pgfplotsset{width=12cm, compat=1.18} 

\usepackage{enumitem}  
\setlist{noitemsep, topsep=0pt, parsep=0pt, partopsep=0pt}

\pgfplotsset{every axis/.append style={line width=0.6pt, tick style={line width=0.5pt}}}

\usepackage{booktabs}
\usepackage{multirow}
\usepackage{array}
\usepackage{tabularx}
\newcolumntype{M}[1]{>{\centering\arraybackslash}m{#1}}

\usepackage[noend]{algorithm2e}
\SetKwFor{ParallelFor}{ParFor}{do in parallel}{end}
\SetKwFor{Parallel}{}{Do in parallel}{end}
\RestyleAlgo{ruled}
\SetKwComment{Comment}{//* }{ *//}

\newtheorem{theorem}{Theorem}
\newtheorem{lemma}[theorem]{Lemma}

\newtheorem{definition}{Definition}

\theoremstyle{definition}  
\newtheorem{example}{Example}
\newtheorem{remark}{Remark}

\numberwithin{equation}{section}

\usepackage[all]{xy}
\usepackage{comment}
\usepackage{caption}
\usepackage{url}
\usepackage{hyperref}

\newcommand{\BibTeX}{\rm B\kern-.05em{\sc i\kern-.025em b}\kern-.08em\TeX}

\begin{document}
 
 \begin{frontmatter}

\title{Efficient  space reduction techniques by optimized  majority rules for the Kemeny aggregation problem}

\author[inst1]{Xuan Kien Phung}
\ead{xuan.kien.phung@umontreal.ca}  
\author[inst1]{Sylvie Hamel}
\ead{sylvie.hamel@umontreal.ca}

\affiliation[inst1]{organization={Département d'informatique et de recherche opérationnelle, Université de Montréal},
            city={Montréal},
            postcode={H3T 1J4}, 
            state={QC},
            country={Canada}}

\begin{abstract}
The Kemeny aggregation problem consists of computing the consensus rankings of an election with respect to the Kemeny-Young voting method. These consensus rankings satisfy various fundamental properties and are the geometric medians of the votes in the election under the Kendall tau distance, which counts the number of pairwise disagreements. The NP-hard  Kemeny aggregation problem admits important applications in various domains such as  operations research, computational social choice, machine learning, and biology but faces a severe bottleneck in practice. In 2020, Milosz and the second author presented an approach to reduce the search space of the problem by solving the relative order of pairs of elements in those consensus rankings.  In this article, we prove an optimized and fully parallelizable extension of this approach. We achieve significantly more space reduction while the parallel design reduces the time complexity from ${O}(n^3)$ to essentially  ${O}(n^2)$ with significant computational speedups in practice. We then show how the constraints built by our approach can be rigorously combined with other methods such as Integer Programming and Finest Condorcet Partitioning to achieve a highly efficient, robust, and scalable solution approach to the Kemeny aggregation problem. Our methods are applicable to the more general constrained Kemeny aggregation ranking problem with immutable  input pairwise constraints. 
Experiments with benchmark against the commercial solver Gurobi are   performed on real and synthetic data. 
\end{abstract}


\end{frontmatter}


\section{Introduction} 
Rank aggregation is a well-studied problem with a wide range of high-stakes applications in many domains. Among several natural distances of the space of all rankings, 
the Kendall-tau distance, which is also the bubble-sort distance between two permutations, is one of the most prominent. It  counts the number of order disagreements between pairs of elements in two permutations. 
In the well-known Kemeny aggregation problem \cite{kemeny, kemeny-snell,young,kendall-tau}, the objective, known as the Kemeny voting rule, is to determine the set of medians that are permutations whose total Kendall-tau distance to the voting profile is minimized. 
Hence,  a median is simply a ranking that maximizes the number of pairwise agreements with the voting profile. The Kemeny voting rule satisfies many desirable properties     from the mathematical and social choice theory point of view such as majority, neutrality, consistency, Condorcet and Smith properties  \cite{young-levenglick,smith}. Moreover, a median is a maximum likelihood estimator of the correct ranking  \cite{kemeny} in the statistical Mallows model \cite{mallows}. Theoretical and practical applications of the Kemeny voting rule vary from decision theory and computational social choice theory \cite{recommender, social, young, young-levenglick}, bioinformatics  \cite{andrieu, brancotte, hamel-bio, 3/4}, recommender systems \cite{Recommender-1}, to 
machine learning and artificial intelligence \cite{3/4,davenport, Kemenywww, soft-condorcet}. Unfortunately, the Kemeny aggregation problem of computing the medians  is   NP-hard \cite{dwork, bachmeier}. Unlocking this computational bottleneck has been  an intensive research area. 
\par 
\textit{Previous works.} 
Existing exact algorithms are computationally expensive and impractical for large-scale instances of the Kemeny aggregation problem. Some authors  have studied exact resolutions using integer linear programming, dynamic programming, branch and bound technique \cite{handbook, bnb1, bnb2}. Other have studied the parametrized complexity and fixed-parameter algorithms of the problem \cite{betzler-jcss, Kenyon, betzler-jcss,karphinski-2010, nishimura-2013}. Most of the literature focuses on  
heuristics \cite{Kemenywww, robin-bnb} and  approximation  algorithms with provable guarantees by innovative and practical approaches. These  range from  greedy methods \cite{cohen-1999, davenport}, linear programming relaxation of the Kemeny aggregation problem  
\cite{ailon, zuylen-deterministic}, and approximations of known exact space reduction techniques  \cite{escobedo} to discrete and smooth approximations of the Kendall-tau distance functions by the Spearman footrule distance or by  sigmoid-type functions 
\cite{dwork, diaconis, soft-condorcet}. An intermediate approach is to obtain exact algorithms to reduce the search space, e.g. by reducing to subproblems of smaller size with the Extended Condorcet Criterion (XCC) \cite{truchon-XCC, escobedo}, with the 3/4 Majority rule \cite{3/4} exploiting non-dirty candidates \cite{betzler-jcss}, or by simplifying the weighted majority graph with the MOT algorithm \cite{hamel-space} and the Unanimity property \cite{always}. 
\par 
\emph{Our contribution}.  The main goal of the paper is to explore and exploit the hidden structures of the  space of all the medians in the context of the Kemeny aggregation problem. We present refined techniques to study various common features shared by all medians ranging from the most basic question \textit{Is x  ranked before y in all the medians?} to more general practical questions such as \textit{What are the top elements in all the medians?}. We then shift the focus to address similar questions concerning a single median.  
\par 
More specifically, our first contribution is  an optimized extension 
of the Major
Order Theorem (MOT) \cite{hamel-space} for the Kemeny aggregation
problem that we call $\alpha$MOT. In essence, MOT and $\alpha$MOT are majority rules which provide sufficient conditions to confirm that the relative rankings of two candidates $x,y$ in \emph{all medians} must respect the preference of the majority of the votes. These easy-to-use and most basic constraints reduce the search space and shed light on the structure of the set of all medians, which can be prohibitively too large and intractable. 
\par 
Our second contribution concerns a relaxation of $\alpha$MOT called $\alpha$MOTe, which guarantees to produce not only constraints obtained by  $\alpha$MOT but also extra valid constraints, always in the form of ordered pairs of candidates, in a certain nonempty subset of medians. Note that $\alpha$MOTe is a majority rule in the sense that it verifies pairs $(x,y)$ with $x$ preferred to $y$ in a majority of votes. Surprisingly,  experimental results show that by using the transitivity property, $\alpha$MOTe constraints can also determine pairs of candidates that we must order \emph{against} the preference of the majority of the votes. Therefore, implications of $\alpha$MOTe go beyond  the scope of majority rules defined above. 
\par 
Our third contribution concerns two major applications. Valid pairwise constraints obtained by $\alpha$MOT and $\alpha$MOTe prove to be very versatile as they can be readily combined with other exact methods such as  Integer Programming and Finest Condorcet Partitioning to  solve large-scale instances of the Kemeny aggregation problem up to a thousand of candidates. 
\par 
For the first application, we incorporate directly the  set $M$ of pairwise constraints obtained from $\alpha$MOT or $\alpha$MOTe to the standard IP formulation of the Kemeny aggregation problem. This is a perfect combination since the standard IP formulation works with pairwise variables $c_{xy}$ and any constraint $(x,y) \in M$ is translated directly to the constraint $c_{xy}=1$. We evaluate the performance of the Gurobi solver on various sets of easy/medium/hard instances with and without the input of  $\alpha$MOT or $\alpha$MOTe constraints. 
\par 
For the second application, we introduce a quadratic time complexity algorithm 
for every  anti-symmetric and anti-reflexive set of constraints  $M$  which provides the unique finest $M$-Condorcet partition. Such a partition is the unique ordered  partition of candidates with the most number of blocks which  respects the constraints $M$ and satisfies the generalized Extended Condorcet Criterion under constraints $M$. This technique allows a  highly desirable and efficient solving strategy: it suffices to solve each block separately then concatenate the solutions of such block to obtain global solutions of the original problem. 
 Our finest $M$-Condorcet partitioning technique seamlessly incorporates 
any set of pairwise constraints $M$ while $\alpha$MOT is applicable to valid sets of constraints satisfied by some median of the election.  At the same time, these methods provide  
a framework to generate constraints and exploit them. 
 
 \usetikzlibrary{positioning, arrows.meta, shadows}

\definecolor{navyblue}{RGB}{31, 78, 121}
\definecolor{lightgrey}{RGB}{245, 247, 250}
\definecolor{bordergrey}{RGB}{180, 190, 201}
\definecolor{textdark}{RGB}{40, 40, 40}

\tikzset{
    pipeline_step/.style={
        rectangle,
        draw=navyblue,
        fill=lightgrey,
        text=textdark,
        line width=1.2pt,
        rounded corners=4pt,
        align=center,
        minimum width=4.5cm,
        minimum height=1.2cm,
        font=\sffamily\small\bfseries,
        inner sep=8pt,
        blur shadow={shadow xshift=1pt, shadow yshift=-1pt, shadow opacity=15}    
    },
    annotation/.style={
        rectangle,
        draw=bordergrey,
        fill=white,
        text=textdark,
        dashed,
        line width=0.8pt,
        rounded corners=2pt,
        align=left,
        minimum width=3cm,
        font=\sffamily\scriptsize,
        inner sep=6pt
    },
    pipeline_arrow/.style={
        ->,
        >=Stealth,
        navyblue,
        line width=1.5pt
    },
    connect_arrow/.style={
        -,
        draw=bordergrey,
        dashed,
        line width=0.8pt
    }
}

\par 
The following Figure~\ref{fig:pipeline} describes an overview of our contributions to the pipeline from preprocessing to solving the Kemeny aggregation problem and its general constrained extension (see the definition~\eqref{e:general} in Section~\ref{s:background}).

\begin{figure}[htbp]
    \centering
    \begin{tikzpicture} 
    [
        node distance=0.5cm and 0.2cm,  
        scale=0.85,                   
        transform shape                
    ]

        \node[pipeline_step] (input)  {Input data (Sec. 2) \\ \normalfont\small (NP Matrix \&   Input  Constraints)};
        
        \node[pipeline_step, below=of input] (preprocess1) {Preprocessing I (Sec. 3-4)\\ \normalfont\small (Output Pairwise Constraints)};
        
        \node[pipeline_step, below=of preprocess1] (preprocess2) {Preprocessing II (Sec. 5)\\ \normalfont\small (Ordered Condorcet Partitions)};
        
        \node[pipeline_step, below=of preprocess2] (solve) {Solving (Sec. 6-7)\\ \normalfont\small (Exact MILP Solvers)};
        
        \node[pipeline_step, below=of solve] (output) {Output  Solutions \\ \normalfont\small (Exact optimal solutions)};

        
               \node[annotation, right=of input] (note_input)  { 
            $\bullet$  {\bfseries\boldmath NP Matrix $[\delta_{xy}]_{x,y \in C}$}  \\
            $\bullet$  {\bfseries\boldmath Input Constraints $W\subseteq C^2$} 
            };

        \node[annotation, right=of preprocess1] (note_preprocess1) {
            $\bullet$ {\bfseries\boldmath $\alpha$MOT (all medians) } \\
            $\bullet$ {\bfseries\boldmath  $\alpha$MOTe (some median) }  
\\
   $\bullet$ {\bfseries\boldmath  $\alpha$MOTe-Inf (some median) }  
}; 

   \node[annotation, right=of preprocess2] (note_preprocess2) {
            $\bullet$ {\bfseries\boldmath  $\alpha$XCC (all medians) }\\
            $\bullet$ {\bfseries\boldmath    $\alpha$XCCe (some median)   }      
            \\
               $\bullet$ {\bfseries\boldmath   $\alpha$XCCe-Inf     (some median) }
              };

        \node[annotation, right=of solve] (note_solve) {  
            $\bullet$ {\bfseries\boldmath Feed output constraints to solvers}\\
            $\bullet$ {\bfseries\boldmath Solve each block of the partition} 
        };

   \node[annotation, right=of output] (note_output) {
            $\bullet$   {\bfseries\boldmath Concatenate sub-solutions}        };

        \draw[pipeline_arrow] (input) -- (preprocess1);
        \draw[pipeline_arrow] (preprocess1) -- (preprocess2);
        \draw[pipeline_arrow] (preprocess2) -- (solve);
        \draw[pipeline_arrow] (solve) -- (output);
        
        \draw[connect_arrow] (input) -- (note_input);
        \draw[connect_arrow] (preprocess1) -- (note_preprocess1);
        \draw[connect_arrow] (preprocess2) -- (note_preprocess2);
        \draw[connect_arrow] (solve) -- (note_solve);
	  \draw[connect_arrow] (output) -- (note_output);

    \end{tikzpicture}
    \caption{Overview of the main contributions of this paper in the preprocessing steps to solve the extended Kemeny problem with input pairwise constraints.}
    \label{fig:pipeline}
\end{figure}

The paper is organized as follows.  Basic notations and background on the Kemeny aggregation problem are given in Section~\ref{s:background} together with a definition of the more general Kemeny aggregation problem with immutable input pairwise constraints. The technical core and algorithmic analysis of $\alpha$MOT are presented in Section~\ref{s:preprocessing-I}. We explore in Section~\ref{s:aMOTe-G1-G2} two rigorous approaches to obtain even more pairwise contraints beyond the limit of $\alpha$MOT namely $\alpha$MOTe and LowInf-Pairs. In Section~\ref{s:preprocessing-II}, we define finest $M$-Condorcet partition and study its fundamental theoretic and algorithmic properties. Section~\ref{s:ip-formulation} presents the augmented IP formulation of the Kemeny aggregation problem by taking into account a set of pairwise constraints.  Experimental results,  discussion, and perspectives are presented in Sections~\ref{s:simulation} to \ref{s:discussion}. See the appendices for the full proofs of the technical results.

\section{Backgrounds and notations}
\label{s:background}

 Following the terminology in social choice theory, an election $(C,V)$ is a finite collection $C=\{c_1, \dots, c_n\}$ of $n=|C|$ candidates (also called alternatives) together with a voting profile $V$ consisting of $m$ votes not necessarily distinct. Here, a ranking or a vote is simply a complete and strict total ordering $\pi \colon c_{\pi(1)}> c_{\pi(2)}> \dots >c_{\pi(n)}$ which we identify with a permutation of elements of $C$. The notation $x>y$ means that $x$ is ranked before $y$.  
Let $S(C)$ be the set of all rankings over the set of candidates $C$. 
The \emph{Kendall-tau distance} $ d_{KT}(\pi, \sigma)$ between two rankings $\pi, \sigma$ of $C$ is: 
\begin{equation}
\label{e:definition-k-wise-distance}
    d_{KT}(\pi, \sigma) = \sum_{\{x,y\} \subset C} \textbf{1}({\mathrm{top}(\pi,{\{x,y\}}) \neq  \mathrm{top}(\sigma,{\{x,y\}})}) \vspace{-0.2cm}
\end{equation}
where $\mathrm{top}(\pi,{\{x,y\}}) \in \{x,y\}$ denotes the highest ranked element in the induced ranking $\pi\vert_{\{x,y\}}$ of $\pi$ on $\{x,y\}$ and the indicator function $\mathbf{1}(X)=1$ if $X$ is true and $\mathbf{1}(X)=0$ if $X$ is false. 
The Kendall-tau distance between a ranking $\pi$ of $C$ and a collection of rankings $A$ of $C$ is defined formally as $
    d_{KT}(\pi,A) = \displaystyle \sum_{\sigma \in A}  d_{KT}(\pi,\sigma)$. A ranking $\pi \in S(C)$ is a \emph{median} of $(C,V)$ if 
$   d_{KT}(\pi,V) = \displaystyle \min_{\sigma \in S(C)} d_{KT} (\sigma, V)$. 
We can associate a weighted majority graph to $(C,V)$ where the set of vertices is $C$ and an arc from a vertex $x$ to another vertex $y$ is weighted by the proportion $a_{xy}$ of the votes in which $x$ is ranked before $y$. 
Equivalently, this weighted majority graph can be presented by the Net Preference (NP) matrix  $[\delta_{xy}]_{x,y \in C}$ (or the Cumulative Ranking matrix) where 
$\delta_{xy} =  ma_{xy} - ma_{yx}$  is the net preference difference between $x$ and $y$.  
\par 
We also consider the more general {Kemeny aggregation problem with immutable input pairwise constraints}. More specifically, let $W\subseteq C^2$ be a set of pairwise constraints   
and let  $(C,V,W)$ denote the corresponding constrained election. We address the constrained optimization problem 
\begin{align}
\label{e:general}
  \min_{\sigma \in S(C)} \quad & d_{KT} (\sigma, V)  \\
\mathrm{s.t.}\quad  
&  x > y \text{ in } \sigma \text{ for all } (x,y) \in W.  \nonumber  
\end{align}
 Optimal solutions of this  problem are called medians of  $(C,V,W)$. We say that $W\subseteq C^2$ is a valid set of constraints of the election $(C,V)$ if there exists a median of the unconstrained election $(C,V)$ that satisfies $W$. In this case, the constrained election $(C,V,W)$ and the unconstrained one $(C,V)$  share the same optimal value but the later may have more optimal solutions. 
 
 \subsection{The Major Order Theorem (MOT)} 
\label{s:mot}
In \cite{hamel-space}, some of the most efficient preprocessing techniques to reduce the search space of the Kemeny aggregation problem are established in the form of the Major Order Theorem (MOT) and its variants. MOT aims to determine pairs of candidates $\{x,y\}$ with $\delta_{xy} > 0$ such that $x$ is ranked before $y$ in every median of the election. More specifically, they defined the interference set   $E_{xy}$ as the collection (with possible repetitions of elements) of candidates  $z$ such that $x>z>y$ in some vote. The multiplicity of $z$ in $E_{xy}$ is the number of votes in which  $x>z>y$.  It was proved in \cite{hamel-space} that if  $\delta_{xy} > |E_{yx} \setminus E_{xy}|$ then $x$ is ranked before   $y$ in every median, that is, $(x,y)$ is a valid constraint in all medians. 
Iterating the above verification by removing from $E_{xy}$ all the candidates which must be ranked before $y$ or after $x$ according to the $i$ previous iterations, we obtain the $(i+1)$-th interference set  $E^{(i+1)}_{xy} \subseteq E^{(i)}_{xy} $. To simplify the notations, we also denote the resulting iterated method proved in \cite{hamel-space} by MOT: if   $\delta_{xy} > |E^{(i)}_{yx} \setminus E^{(i)}_{xy}|$ for some $i \geq 0$, then $x$ is ranked before $y$ in every median. Each  MOT verification for a pair $(x,y)$ requires $O(n)$ time. 
\par 
As formulated in the original paper, the input to MOT must be an election rather than merely a weighted tournament since the computation of the interference sets relies on the underlying vote profiles. 
 However, we only need to compute the cardinality $|E^{(i)}_{yx} \setminus E^{(i)}_{xy}|$ to check the conditions $\delta_{xy} > |E^{(i)}_{yx} \setminus E^{(i)}_{xy}|$. It turns out that $ |E^{(i)}_{yx} \setminus E^{(i)}_{xy}|$ can be computed directly from the weight vector $[b_{xy}]_{x,y \in C}$ of the tournament without any access required to actual votes. Hence, MOT works as well on weighted tournaments. 
 
\subsection{Extended Condorcet Criterion (XCC)} 
\label{s:xcc-acp}

The  \emph{Condorcet criterion}  states that  if we have a partition $C=X\cup Y$ in which $\delta_{xy}>0$ for all $x\in X$ and $y \in Y$ then every median $\pi$ respects such a partition, that is, $x$ is ranked before $y$ in the ranking $\pi$ for all $x\in X$ and $y \in Y$ \cite{condorcet}. 
The \emph{Extended Condorcet Criterion} (XCC) discovered in \cite{truchon-XCC} generalizes the Condorcet criterion 
 and states that given an election and a partition $C=\bigcup_{i=1}^k X_i$ such that $\delta_{x_ix_j}>0$ for all $x_i\in X_i$ and $x_j \in X_j$ with $i<j$, then in every median $\pi$ of the election, every element of $X_i$ is ranked before every element of $X_j$  for all $i<j$. 
 
We call such a partition $C=\bigcup_{i=1}^k X_i$ an {XCC partition} and $X_1,\dots,X_k$ its blocks.  Following \cite{escobedo}, a finest XCC partition is one with the maximal number of blocks. An algorithm with time complexity $O(n^2)$ was obtained in \cite{escobedo} to produce the finest XCC partition.

\section{Preprocessing I: $\pmb \alpha$MOT constraints} 
\label{s:preprocessing-I}
For the preprocessing step I, we begin by describing the foundation $\alpha$MOT in Section~\ref{s:aMOT},  an optimized extension of MOT, to capture more pairwise constraints for all medians. In Section~\ref{s:single-alpha-mot}, we   analyze  the core algorithm $\alpha$MOT-SinglePair verifying a single pairwise constraint at a time.  In Section~\ref{s:algo-iterated-ab-mot}, Algorithm~\ref{alg:Iterated-abMOT} for $\alpha$MOT is obtained by calling sequentially $\alpha$MOT-SinglePair. We then discuss a  parallel algorithm for $\alpha$MOT in Section~\ref{s:parallel}. Several variants of $\alpha$MOT to obtain even more constraints will be discussed  in Section~\ref{s:aMOTe-G1-G2} to complete the preprocessing I step.  

\subsection{Statement and idea of $ \alpha$MOT}
\label{s:aMOT}

For the notation, let $U\subseteq C^2$ be a set of constraints consisting of ordered pairs of candidates.  We denote by $\overline{U}  \subseteq C^2$ the \emph{transitive closure} of $U$, i.e., the smallest transitive subset of $C^2$ containing $U$. 
 Given $x\in C$, we define the set $  L_{x}(U)  $ of  elements ranked to the left of $x$ and similarly the set $ R_{x}(U)  $ of  elements ranked to right of $x$ according to the set of constraints $U$ by 
    \begin{align*} 
    L_{x}(U)  =\{z \colon (z,x)  \in {U} \}, 
  \quad 
 R_{x}(U) & =\{z \colon (x,z) \in {U} \}. 
 \end{align*} 
 
\par 
Our first main result, $\alpha$MOT, aims to verify  for each pair $(x,y)$ of candidates whether $x>y$ in every median. To achieve this, we compare a partial ranking $yZx$ to a weighted combination of $xyZ$ and $Zxy$ where $Z$ is a  collection of possible elements  that can be ranked between $y$ and $x$ in that order in some median, taking into account all  known  constraints. Then either $xyZ$ or $Zxy$ is more optimal than $yZx$ provided that the preference of $x$ over $y$ is strong enough even in the presence of any collection and combination of elements in the interfering set  (see eq. \eqref{e:alpha-beta-2-mot-condition}   for the precise condition). Each pair that is successfully solved will help reducing the size of the interfering set of other pairs yet to be solved. We iterate the above   process until saturation. 
 
\begin{theorem}[$\alpha$MOT]
\label{t:alpha-beta-2MOT-3.0} 
Let $(C,V)$ be an election together with a valid set of constraints $W\subseteq C^2$.  
Let $W_{-1}=W$ and 
for every $k \geq 0$, we define $W_k$  inductively as the set of ordered pairs $(x,y)$ of distinct candidates such that
\begin{align}
\label{e:alpha-beta-2-mot-condition}
     \delta_{xy}  > \min_{\alpha, \beta\geq 0,\,  \alpha+\beta=1} \sum_{z\in\,  Z_k(x,y)}  \max\left(0, \alpha \delta_{yz} + \beta \delta_{zx} \right)
\end{align}
where the interference set $Z_k(x,y)$ is defined by 
\begin{align} 
\label{e:z-k}
Z_k(x,y)= C\setminus \left( L_{y}\left(\overline{W}_{k-1}\right) \cup R_{x}\left(\overline{W}_{k-1}\right) \cup \{x,y\}\right).
\end{align} 
Suppose that  $(x,y) \in \overline{W}_{k}$ for  some   $k \geq -1$. Then $x$ is ranked before $y$ in every median of the constrained election $(C,V,W)$.  
\end{theorem}

   \begin{proof} 
We proceed by induction on $k \geq -1$. The case $k=-1$ is trivial since $W_{-1}=W$. Assume that the conclusion holds for some   $k \geq -1$ and let $(x,y) \in W_{k+1}$. Suppose, on the contrary, that  $\pi \colon L>y>Z>x>R$ is a median  where $L,Z,R$ are ordered sets of candidates. 
Let  $\sigma_1 \colon L>Z>x>y>R$ and  $\sigma_2\colon L>x>y>Z>R$.  Let $
 \Delta_i= d_{KT} (\pi, V) - d_{KT}(\sigma_i, V)$ where $i\in \{1,2\}$. Let $\alpha, \beta \geq 0$ with $\alpha+\beta=1$ such that $\sum_{z\in Z_k(x,y)}  \max\left(0, \alpha \delta_{yz} + \beta \delta_{zx} \right)$ is minimized (please refer to Section~3.2 for the existence of $\alpha$ and $\beta$).  
 By the induction hypothesis, 
$L_{y}\left(\overline{W}_{k}\right) \subseteq L$ and $R_{x}\left(\overline{W}_k\right) \subseteq R$ so   
$ 
Z \subseteq Z_{k+1}(x,y) = C\setminus \left(L_{y}\left(\overline{W}_k\right) \cup R_{x}\left(\overline{W}_k\right) \cup \{x,y\} \right)$. 
Since $(x,y) \in W_{k+1}$ and $Z\subseteq Z_{k+1}(x,y)$, we can estimate   
$ \alpha \Delta_1 +  \beta \Delta_2  = \alpha\left(  \delta_{xy} -   \sum_{z \in Z}  \delta_{yz} \right) + \beta \left(  \delta_{xy} -   \sum_{z \in Z}\delta_{zx} \right)    = (\alpha+\beta) \delta_{xy} -    \sum_{z \in Z} \left(\alpha \delta_{yz} + \beta \delta_{zx} \right) =   \delta_{xy} -    \sum_{z \in Z} \left(\alpha  \delta_{yz} + \beta \delta_{zx} \right) 
\geq \delta_{xy} -    \sum_{z \in Z_{k+1}(x,y)} \max \left(0, \alpha \delta_{yz} + \beta \delta_{zx} \right) >0$. 
Hence, $\Delta_j >0$ for some $j \in \{1,2\}$ and $\sigma_j$ has a strictly lower Kemeny score  than $\pi$, which contradicts the choice of $\pi$. Hence, $x$ must be ranked before $y$ in every  median. By transitivity, we conclude that  $x$ is ranked before $y$ in every  median if $(x,y) \in \overline{W}_{k+1}$.
\end{proof}

\par 
 
\begin{remark}   
 The  iterated version of MOT  can be obtained as a particular case of Theorem~\ref{t:alpha-beta-2MOT-3.0} if we fix $W=\varnothing$ and  $\alpha=\beta=0.5$ in the  condition~\eqref{e:alpha-beta-2-mot-condition} which results in  a stronger condition 
 $ \delta_{xy}> 0.5\sum_{z\in\,  Z_k(x,y)}  \max\left(0,  \delta_{yz} + \delta_{zx} \right)$. 
In other words, $\alpha$MOT is  a certain optimized weighted version of MOT.
\end{remark}  

\begin{remark} 
Ideally, we want to  check for \emph{all subsets} $Z\subseteq Z_k(x,y)$ that 
  $2\delta_{xy} -    \sum_{z \in Z} \left(  \delta_{yz} +  \delta_{zx} \right) >0$ but a direct computation is not feasible as there is an exponential number of such subsets $Z$. However, it suffices to check the slighter stronger but tractable condition \eqref{e:alpha-beta-2-mot-condition}  (see Section~\ref{s:algo-iterated-ab-mot}). 
 \end{remark} 
\par 
We describe below a running toy example illustrating our methods. 
\begin{example}
\label{ex:toy-example-1}
    Let $C=\{1,2,3,4,5,6,7,8\}$ be a set of 8 candidates and let $V=\{\pi_1, \pi_2, \pi_3, \pi_4\}$ be a  voting profile over $C$ consisting of 4 votes  $\pi_1 \colon 3>7>1>4>2>5>8>6$,  
 $\pi_2\colon 
        2>5>8>1>3>4>7>6$, 
 $\pi_3\colon 3>6>8>7>4>1>2>5$, and 
 $\pi_4\colon 5>2>8>3>1>7>4>6$. 
All the medians of the above election have Kemeny distance 35 to $V$ which are $\pi   \colon 2>5>8>3>1>7>4>6$,
  $\sigma_1\colon 2>5>8>3>7>1>4>6$,
 $\sigma_2\colon 3> 2>5>8>7>1>4>6$,
 $\sigma_3\colon 3> 2>5>8>1>7>4>6$,
 $\sigma_4\colon  2>5>3>8>7>1>4>6$,
 $\sigma_5\colon  2>5>3>8>1>7>4>6$,
  $\sigma_6\colon 2>3>5>8>7>1>4>6$, and 
 $\sigma_7\colon 2>3>5>8>1>7>4>6$. 
A   ranking of $C$ is determined by a list of pairwise comparisons of the elements in each of the  $\binom{8}{2}=28$ pairs. We identify such a pairwise comparison with an ordered pair where $(i,j)$ means that $i$ is ranked before $j$. 
In our example, MOT determines in total 13 ordered pairs in all medians which are $(1,4)$, $(1,6)$, $(3,1)$, $(3,4)$, $(3,6)$, $(3,7)$, $(4,6)$, $(5,6)$, $(7,6)$, and $(8,6)$. Adding to these, $\alpha$MOT solves an extra 11 ordered pairs in all medians namely $(2,1)$, $(2,4)$, $(2,5)$, $(2,6)$, $(2,7)$, $(2,8)$, $(5,1)$, $(3,4)$, $(5,7)$, $(5,8)$, $(8,1)$, and $(8,7)$. 
 \end{example}

\subsection{Computing $\alpha$MOT constraints} 
\label{s:single-alpha-mot}

Given Theorem~\ref{t:alpha-beta-2MOT-3.0}, our main task reduces to efficiently computing the right-hand side (RHS) of the condition  \eqref{e:alpha-beta-2-mot-condition}.  For this, let us rewrite the RHS of \eqref{e:alpha-beta-2-mot-condition} as a function $F_k\colon [0,1] \to \mathbb{R}$: 
$$
F_k(\alpha)=   \sum_{z\in Z_k}  \max\left(0, \alpha \delta_{yz} + (1-\alpha) \delta_{zx} \right)=\sum_{z\in Z_k} f_{x,y,z}(\alpha). 
$$
where $f_{x,y,z}(\alpha)= \max\left(0, \alpha \delta_{yz} + (1-\alpha) \delta_{zx} \right)$ and $Z_k=Z_k(x,y)$ is defined by the relation~\eqref{e:z-k}.  
First, observe that $F_k$ is a continuous piecewise linear function as a sum of piecewise linear functions $f_{x,y,z}$ on $[0,1]$. Second,  we consider the critical points $0=t_1, \dots, t_r=1$ of $F_k$ on the interval $[0,1]$ which are points of discontinuity of $F_k$ plus the two boundary points $0$ and $1$ of the interval $[0,1]$. These points form a subset of the union $T_k$ of critical points of the functions $f_{x,y,z}$ for all $z \in Z_k$. Note that 
\begin{itemize}
    \item if $\delta_{yz}\delta_{zx} \geq 0$, the critical points of $f_{x,y,z}$ are $0$ and $1$; 
    \item if  $\delta_{yz}\delta_{zx} < 0$, the critical points of $f_{x,y,z}$ are  $0$, $1$, and $\displaystyle \frac{\delta_{zx}}{\delta_{zx}-\delta_{yz}}\in (0,1)$.
    \end{itemize} 
    Hence, $r \leq |T_k|\leq   |Z_k| +2 \leq |C|-2+2=|C|=n$. 
    Finally, we can thus efficiently compute in worst-case quadratic time $O(n^2)$   
\begin{align}
 \label{e:F-alpha-beta} 
  \min_{\alpha\in [0,1]}   F_k(\alpha)=\min_{1 \leq i\leq r} F_k(t_i)=\min_{t\in T_k} F_k(t). 
\end{align}

\par 
We summarize the above core technical discussion to compute the $\alpha$MOT constraints in the following Algorithm called $\alpha$MOT-SinglePair.

\begin{algorithm}[ht]
\caption{$\alpha$MOT-SinglePair}\label{alg:mot-pair}
\footnotesize              \linespread{1.0}\selectfont 
\DontPrintSemicolon 
\KwData{candidates $x,y\in C$, matrix $[\delta_{uv}]_{u,v\in C}$, constraints $W\subseteq C^2$}
\KwResult{ $\displaystyle \mathrm{RHS} \xleftarrow{\mathrm{Eq. } \eqref{e:alpha-beta-2-mot-condition}} \min_{\alpha, \beta\geq 0,\,  \alpha+\beta=1} \sum_{z\in Z}  \max\left(0, \alpha \delta_{yz} + \beta \delta_{zx} \right)$}

$Z \gets C\setminus \left( L_{y}\left({W}\right) \cup R_{x}\left({W} \right) \cup \{x,y\}\right)$ \; 
 $\mathrm{RHS}\gets\displaystyle \min_{\alpha, \beta\geq 0,\,  \alpha+\beta=1} \sum_{z\in Z}  \max\left(0, \alpha \delta_{yz} + \beta \delta_{zx} \right)$ \; \Comment*[r]{\small use the formula \eqref{e:F-alpha-beta} with $Z_k(x,y)=Z$} 

\KwRet{$\mathrm{RHS}$} 
\end{algorithm}

 \begin{theorem}
In Algorithm $\alpha \mathrm{MOT}\text{-}\mathrm{SinglePair}$, the time complexity for the verification of a pair of candidates $(x,y)$ is  $O(|Z_k|^2) $. Here,  the set $Z_k=Z_k(x,y)$ is given by the formula \eqref{e:z-k} and satisfies $|Z_k|\leq n-2$. 
\end{theorem}
 
As $|Z_k|\leq n$, the worst-case complexity is $O(n^2)$. However, when more constraints are solved,  $Z_k(x,y)$ becomes smaller very quickly. 
This observation also motivates Algorithm LowInf-Pairs in Section~\ref{s:low-inf}. 
\par 
Another critical feature of conditions~\eqref{e:alpha-beta-2-mot-condition} and~\eqref{e:F-alpha-beta} is that they can be expressed in an equivalent form using exclusively integral quantities. This holds true even when the non-trivial evaluation points $t \in T_k$ of $F_k(t)$ in~\eqref{e:F-alpha-beta} are highly fractional. This property guarantees the exact nature of our algorithms and ensures the deterministic validity of the $\alpha$MOT and $\alpha$MOTe constraints, thereby eliminating any potential floating-point rounding errors during execution. For comparison, numerical rounding issues are intrinsic to many commercial mathematical programming solvers, such as Gurobi, and often represent a non-negotiable tradeoff between computational speed and numerical precision. Consequently, such solvers frequently converge to slightly suboptimal solutions when handling highly fractional or ill-conditioned complex instances. We formalize this feature of $\alpha$MOT as follows. 
 
\begin{lemma}(Integer-only equivalent conditions)
Fix $k \in \mathbb{N}^*$. Let $T_k = \{t_1, \dots, t_r\}$ with $\displaystyle t_i = \frac{a_i}{b_i}$, where $a_i, b_i \in \mathbb{N}^*$ and $b_i \neq 0$. Let $\displaystyle f_j = \min_{1 \leq i \leq j} F_k(t_i)$. Then $\displaystyle f_r = \min_{\alpha \in [0,1]} F_k(\alpha)$, and we can write $\displaystyle f_j = \frac{\mathcal{N}_j}{\mathcal{D}_j}$, where $\mathcal{N}_j, \mathcal{D}_j \in \mathbb{N}$ can be computed recursively on $j$ using only integer operations as follows:
\begin{align} 
\label{e:integer-cross-product-1}
N_j & = \sum_{z \in Z_k(x,y)} \max\left(0, a_j \delta_{yz} + (b_j - a_j) \delta_{zx} \right), \\
\label{e:integer-cross-product-2}
(\mathcal{N}_{j+1}, \mathcal{D}_{j+1}) & =  
\begin{cases}
(N_{j+1}, b_{j+1}) & \mathrm{if\ } N_{j+1} \mathcal{D}_j < \mathcal{N}_j b_{j+1}, \\ 
(\mathcal{N}_j, \mathcal{D}_j) & \mathrm{otherwise.}   
\end{cases} 
\end{align} 
\end{lemma}

\subsection{Sequential algorithm for   $ \alpha$MOT} 
\label{s:algo-iterated-ab-mot}

Based on the pairwise verification  $\alpha\text{MOT-SinglePair}(x,y,[\delta_{uv}],W)$  for each pair $(x,y)\in C^2$, Algorithm \ref{alg:Iterated-abMOT} below computes sequentially and iteratively the constraints $\alpha$MOT. 
   In general, the number of new pairs solved is decreasing after each iteration with the highest concentration in the first two. In other words, the cost of solving a new pair is increasing   after each round.

 \begin{algorithm}[ht]
\caption{Sequential $\alpha$MOT}\label{alg:Iterated-abMOT}
\footnotesize               \linespread{1.0}\selectfont 
\DontPrintSemicolon 
\KwData{candidates $C$, matrix $[\delta_{uv}]_{u,v\in C}$, constraints $W\subseteq C^2$}
\KwResult{a set of constraints $M\subseteq C^2$ containing $W$, and  a set $E\subseteq C^2$} 
$E \gets \varnothing$; \, $\mathrm{len} \gets -1$; \, $M \gets W$\; 
\While{$\mathrm{len} < |M|$}{
  $\mathrm{len} \gets |M|$\;
  {\For{$(x,y) \in C^2$ : $\delta_{xy}>0$, $(x,y)\notin M$}{$\mathrm{RHS} \gets \alpha\text{MOT-SinglePair}(x,y,[\delta_{uv}],M)$  
   
    \If{$\delta_{xy}>\mathrm{RHS}$}{append  $(x,y)$ to $M$}
    \If{$\delta_{xy} =  \mathrm{RHS}$}{append  $(x,y)$ to $E$}
    }

    $M\gets \overline{M}$\; 
  }
}
\KwRet{$(M,E)$} 
\end{algorithm}

\subsection{Parallel algorithm for $ \alpha$MOT} 
\label{s:parallel}

The  verifications $\alpha\text{MOT-SinglePair}(x,y,[\delta_{uv}], W) < \delta_{xy}$ for individual pairs of candidates can be executed simultaneously and do not require strict sequential dependency. This independence allows for scalable parallel algorithms designed to compute the complete set of $\alpha$MOT constraints. 
\par 
An efficient architectural solution, which we adopt in our experimental framework, divides the candidate pairs across dedicated available processing cores. A defining feature of this  approach is the utilization of a separate, isolated core tasked exclusively with continuously collecting newly solved pairwise constraints from other workers and continuously executing the transitivity closure. 
This mechanism is formalized in Algorithm~\ref{alg:parallel-abMOT}, which provides a simplified yet highly efficient implementation of $\alpha$MOT in a parallel fashion. Figure~\ref{fig:parallel_computing_architecture} also gives an overview of our implementation.  

  \begin{algorithm}[ht]
\caption{Parallel $\alpha$MOT}\
\label{alg:parallel-abMOT}
\footnotesize          
    \linespread{1.0}\selectfont 
\DontPrintSemicolon 
\KwData{candidates $C$, NP matrix $[\delta_{uv}]_{u,v\in C}$, initial constraints $W\subseteq C^2$}
\KwResult{a set of constraints $M\subseteq C^2$ containing $W$, and a set $E\subseteq C^2$} 
$E \gets \varnothing$ \Comment*[r]{ Shared equality-type constraints buffer} 
$M \gets {W}$ \Comment*[r]{ Shared global constraint matrix} 
$R \gets \{(x,y)\in C^2:\delta_{xy}>0\}\setminus M$ \Comment*[r]{Shared pool of remaining pairs}
$\mathrm{newpairs}\gets \text{True}$ \Comment*[r]{ Shared global control flag} 

\Parallel{}{
    \While{$\mathrm{newpairs}$}{ 
        Remove all resolved pairs $(x,y)\in \overline{M}\setminus M$ from $R$\;
        $M\gets \overline{M}$\;
    }  
    \While{$\mathrm{newpairs}$}{
        $\mathrm{temp1}\gets \mathrm{False}$; \,   $\mathrm{temp2}\gets \mathrm{len}(M)$\;
        \ParallelFor{$(x,y) \in R$}{
            $\mathrm{RHS} \xleftarrow{\mathrm{Eq. } \eqref{e:alpha-beta-2-mot-condition}} \alpha\text{MOT-SinglePair}(x,y,[\delta_{uv}]_{u,v\in C},M)$\;
            \If{$\delta_{xy}>\mathrm{RHS}$}{
                Add $(x,y)$ to $M$ and remove $(x,y)$ from $R$\; 
                $\mathrm{temp1}\gets \mathrm{True}$\;
            }
            \If{$\delta_{xy} = \mathrm{RHS}$}{
                Append $(x,y)$ to $E$\;
            }
        }
        \If {$\mathrm{temp1}=\mathrm{False}$ $\&$ $\mathrm{temp2}=\mathrm{len}(M)$}{
            $\mathrm{newpairs}\gets\mathrm{False}$\;
        }
    }
}
\KwRet{$(M,E)$}\;
\end{algorithm}
 
 \begin{figure}[htbp]
\centering
\begin{tikzpicture}[
     node distance=0.5cm and 1.1cm,
    >=Stealth,
     blockD/.style={rectangle, draw=navyblue, fill=lightgrey, minimum width=4.9cm, minimum height=2cm, align=center, rounded corners=3px, font=\sffamily \footnotesize, line width=0.8pt},
    blockBC/.style={rectangle, draw=orange!80!black, fill=orange!10, minimum width=3.2cm, minimum height=1.6cm, align=center, rounded corners=3px, font=\sffamily \footnotesize, line width=1.2pt},
    coreStyle/.style={rectangle, draw=navyblue, fill=white, minimum width=1.8cm, minimum height=0.45cm, align=center, rounded corners=2px, font=\sffamily \footnotesize}
]

 
\node[draw=navyblue, fill=lightgrey, line width=1.2pt, rounded corners=4pt, minimum width=4.9cm, minimum height=2cm,  font=\sffamily\footnotesize\bfseries] (D) {};

\node[coreStyle, yshift=-0.5cm] (D0) at (D.center) {Core 0};
\node[font=\sffamily\bfseries\footnotesize, align=center, above=0.07cm of D0] (DLabel) {\textbf{Transitivity closure worker} \\ Indep. Core (Trans-{C})};

\node[draw=navyblue, fill=lightgrey, line width=1.2pt, rounded corners=4px, below=0.35cm of D, minimum width=4.9cm, minimum height=3.8cm,  font=\sffamily\footnotesize\bfseries] (E) {};

\node[coreStyle, yshift=0.36cm] (C1) at (E.center) {Core 1};
\node[coreStyle, yshift=-0.34cm] (C2) at (E.center) {Core 2};
\node[font=\boldmath\footnotesize, yshift=-0.73cm] (Cdots) at (E.center) {$\vdots$};
\node[coreStyle, yshift=-1.35cm] (C7) at (E.center) {Core $\mathcal{H}-1$};

\node[font=\sffamily\bfseries\footnotesize, align=center, above=0.12cm of C1] (ELabel) {Parallel workers computing \\ $\pmb \alpha$MOT-SinglePair};

\node[blockBC, right=1.2cm of D, yshift=-0.4cm] (B) {\textbf{Contraint matrix} \\ Shared (R/W)};

\node[blockBC, below=0.35cm of B] (C) {\textbf{Interference matrix} \\ Shared (R/W)};

\node[blockBC, below=0.35cm of C] (A) {\textbf{Net preference matrix} \\ Shared (R-Only)};

\tikzset{curvedArrow/.style={<->, >= {Triangle[length=6pt, width=4pt]}, line width=1pt, draw=orange!80!black}}
\tikzset{curvedArrowLeft/.style={<-, >= {Triangle[length=6pt, width=4pt]}, line width=1pt, draw=orange!80!black}}

\draw[curvedArrow] (D.east) .. controls +(0.5,0) and +(-0.5,0) .. ($(B.west)!0.2!(B.north west)$);

\draw[curvedArrow] ($(E.east)!0.25!(E.north east)$) .. controls +(0.5,0) and +(-0.5,0) .. ($(B.west)!0.3!(B.south west)$);

\draw[curvedArrow] (E.east) .. controls +(0.5,0) and +(-0.5,0) .. (C.west);

\draw[curvedArrowLeft] ($(E.east)!0.2!(E.south east)$) .. controls +(0.5,0) and +(-0.5,0) .. (A.west);

\end{tikzpicture}
\caption{Parallel computing architecture for $ \alpha$MOT.}
\label{fig:parallel_computing_architecture}
\end{figure}

We dedicate a separate background process (Trans-C), independent of the main worker loops, to repeatedly compute the transitive closure $M \gets \overline{M}$ of the shared constraint set $M$. To eliminate scheduling interference and lock contention on shared memory variables, we enforce strict partitioning over the processor architecture by pinning the Trans-C process to physical core $p_0$, while the primary $\alpha$MOT computational tasks are distributed across the remaining cores $\{p_1, \dots, p_{\mathcal{H}-1}\}$. This asymmetric spatial isolation prevents cache eviction, minimizes memory access latency, and achieves near-linear speedups across the parallel execution pipeline. Concurrently, Algorithm~\ref{alg:parallel-abMOT} distributes the evaluation tasks for pairs $(x,y)$ from a shared pool of remaining unsolved pairs $R \subset C^2 \setminus M$ using $\alpha\text{MOT-SinglePair}$. In practice, the algorithm is executed from scratch with $W = \varnothing$ unless a set of immutable pairwise constraints $W \subseteq C^2$ is imposed.

 \section{How to obtain more constraints?}
 \label{s:aMOTe-G1-G2}
We introduce two rigorous algorithmic extensions designed to maximize the yield of pairwise constraints from $\alpha$MOT. The first variant relaxes the structural requirement by operating on a single median rather than the entire set of medians (see $\alpha$MOTe in Section~\ref{s:one-median-amote}). The second algorithm  achieves denser constraint sets by exploiting  previously computed pairwise constraints at a minor computational extra cost (see Algorithm~\ref{alg:low-inf} in Section~\ref{s:low-inf}).

\subsection{Focus on one median: $ \alpha$MOT$\text{e}$}
\label{s:one-median-amote}
 We can relax the strict inequality in the condition \eqref{e:alpha-beta-2-mot-condition} to obtain more constraints. However, two issues arise. First, when a pair $(x,y)$ realizes the equality in Condition \eqref{e:alpha-beta-2-mot-condition},  $x$ can have a better ranking than  $y$ in only \emph{some medians} instead of  \emph{all medians}. 
 Second, such equality-type constraints may altogether exclude all the medians and thus lead to suboptimal solutions. 
  While the first issue is unavoidable, the following result provides a solution to the second issue and allows us to achieve more valid constraints.  

 \begin{theorem}[$\alpha$MOT{e}]
 \label{alg:abmote}
Let $E$ be the set of ordered pairs realizing the equality in  Condition~\eqref{e:alpha-beta-2-mot-condition} for at least one  $k$. Then for each  $S \subseteq C$, there exists a median satisfying the set of constraints $E(S)=\{(x,y)\in E\colon x\in S, \, y \notin S\}$. 
\end{theorem}

 Let  $S\subseteq C$ be a set maximizing $|E(S)|$ and let  $\alpha$MOTe be the extension of $\alpha$MOT that incorporates  all the constraints in $E(S)$. The corresponding Algorithm~\ref{alg:Iterated-abMOTe} computing  $\alpha$MOTe is given  below.

\begin{algorithm}[ht]
\caption{ $\alpha$MOTe}\label{alg:Iterated-abMOTe}
\label{alg:abMOTe}
\footnotesize           
    \linespread{1.0}\selectfont 
\DontPrintSemicolon 
\KwData{candidates $C$, matrix $[\delta_{xy}]_{x,y \in C^2}$, constraints $W \subseteq C^2$}
\KwResult{a new set of constraints $M\subseteq C^2$ containing $W$}  
$(M,E) \gets \alpha\text{MOT}(C,[\delta_{xy}],W)$\; 
Compute  $S\subseteq C$ which maximizes $|E(S)|$\;  
$F(S) \gets \{(x,y)\in E\colon x\in S, \, y \notin S\}$\; 
$W \gets \overline{M \cup F(S)}$\; 
$M \gets \alpha$MOT$(C,[\delta_{xy}],W)$ \; 
\KwRet{$M$} 
\end{algorithm}
   
Computing a subset $S\subseteq C$ maximizing $|E(S)|$ is an instance of the Directed Maximum Cut Problem. For this part, we use Integer Programming whose running time is observed to be negligible to the rest of Algorithm~\ref{alg:abMOTe}. 

\begin{example}
    \label{ex:toy-example-1-continued} For Example~\ref{ex:toy-example-1}, $\alpha$MOTe solves 4 additional constraints $(1,7)$, $(2,3)$, $(5,3)$, $(8,3)$ (see Figure~\ref{fig:pairs-solved-toy-example-1}) thus provides a complete set of 28 ordered pairs which determines the median $\pi\colon 2>5>8>3>1>7>4>6$.  
\end{example}

\begin{figure}[H] 
\centering 
\begin{tikzpicture}[scale=0.7, transform shape]
 
  \definecolor{navyblue}{RGB}{31, 78, 121}
\definecolor{lightgrey}{RGB}{245, 247, 250}
\definecolor{textdark}{RGB}{40, 50, 60}      
\definecolor{terracotta}{RGB}{198, 102, 87}   
\definecolor{sagegreen}{RGB}{60, 130, 95}     

\tikzset{
    dep/.style={
        circle,
        minimum size=0.7cm,
        fill=lightgrey,
        draw=navyblue,
        text=textdark,
        line width=1.2pt,
        font=\sffamily\bfseries,
        drop shadow={blur shadow, shadow xshift=1pt, shadow yshift=-1pt, shadow opacity=15}
    },
    mot_arrow/.style={
        ->,
        >={Triangle[length=6pt, width=4pt]},
        draw=textdark!85,
        line width=1.2pt
    },
    alpha_arrow/.style={
        ->,
        >={Triangle[length=6pt, width=4pt]},
        draw=sagegreen,
        line width=1.3pt,
        dash pattern=on 2pt off 2pt 
    },
    green_arrow/.style={
        ->,
        >={Triangle[length=6pt, width=4pt]},
        draw=terracotta,
        line width=1.3pt,
        dash pattern=on 2pt off 2pt  
            }
}
\node[dep] (s) at (-2,-1.5) {1};
\node[dep] (t) at (0,0) {2};
\node[dep] (u) at (2.5,0) {3};
\node[dep] (v) at (4.5,-1.5) {4};
\node[dep] (w) at (4.5,-3.5) {5};
\node[dep] (x) at (2.5,-5) {6};
\node[dep] (y) at (0,-5) {7};
\node[dep] (z) at (-2,-3.5) {8};

\draw[mot_arrow] (s) to (v); 
\draw[mot_arrow] (s) to (x); 
  \draw[mot_arrow] (u) to (s); 
\draw[mot_arrow] (u) to (v); 
\draw[mot_arrow] (u) to (x);
\draw[mot_arrow] (u) to (y); 
\draw[mot_arrow] (v) to (x); 
\draw[mot_arrow] (w) to (x); 
\draw[mot_arrow] (y) to (v); 
\draw[mot_arrow] (y) to (x); 
\draw[mot_arrow] (z) to (x); 

 \draw[alpha_arrow] (t) to (s);
\draw[alpha_arrow] (t) to (v);
\draw[alpha_arrow] (t) to (w);
\draw[alpha_arrow] (t) to (x);
\draw[alpha_arrow] (t) to (y);
\draw[alpha_arrow] (t) to (z);
\draw[alpha_arrow] (w) to (s);
\draw[alpha_arrow] (w) to (v);
\draw[alpha_arrow] (w) to (y);
\draw[alpha_arrow] (w) to (z);
\draw[alpha_arrow] (z) to (s);
\draw[alpha_arrow] (z) to (v);
\draw[alpha_arrow] (z) to (y);

 \draw[green_arrow] (s) to (y);
\draw[green_arrow] (t) to (u);
\draw[green_arrow] (w) to (u);
\draw[green_arrow] (z) to (u);
\end{tikzpicture}

 \caption{Digraph showing all 10 ordered pairs solved by MOT (gray arrows) and the extra 12 ordered pairs solved by $\alpha$MOT (dashed green arrows) in all medians of the election given in Example~\ref{ex:toy-example-1}. The extra 4 ordered pairs solved by $\alpha$MOTe are dashed red arrows. }
\label{fig:pairs-solved-toy-example-1} 
\end{figure}

 \subsection{Solving  pairs with low interference} 
 \label{s:low-inf} 
Our goal in this section is to obtain even more constraints by further exploiting the output of $\alpha$MOT and $\alpha$MOTe. The main idea is that given the set of constraints solved by $\alpha$MOT and $\alpha$MOTe, we can easily identify pairs with a bounded small number of possible elements lying between them in the medians. Pairs with this property are called pairs with low interference. We then only need to solve a small Kemeny instance to verify if such a pair must respect the preference of a majority of the votes. More specifically, we define: 
 \begin{definition} 
 Given $M\subseteq C^2$, the $M$-interference set of an ordered pair $(x,y)\in C^2$ is defined by  
$  I_M(x,y) = \{z\in C\setminus \{x,y\}\colon   (z,x)\notin M, \, (y,z)\notin M\}$. 
 \end{definition}
\noindent  Generalizing the set $Z_k(x,y)$ in Theorem~\ref{t:alpha-beta-2MOT-3.0}, the set $I_M(x,y)$ is the set of all candidates that can appear between $y$ and $x$ in that order among all possible rankings that satisfy $M$. 
Suppose that $(x,y)$ is a pair with low $M$-interference $|I_M(x,y)|$. If $y$ is not the winner in any median of the election restricted to every union of $\{x,y\}$ with a subset of $ I_M(x,y)$, then  $y$ cannot be ranked before $x$ in any median. In this case, we can  safely add the valid constraint $(x,y)$ to $M$. If $M$ only contains    constraints satisfied by some median $\pi$ then such a constraint $(x,y)$ is also a valid constraint satisfied by $\pi$.   We  formalize in Lemma~\ref{l:low-interference} below the method to obtain more constraints targeting pairs  with low interference.  The resulting algorithm is LowInf-Pairs (Algorithm~\ref{alg:low-inf}).  
 
 \begin{lemma} 
 \label{l:low-interference}
 Let $M\subseteq C^2$ be a set of constraints satisfied by some median $\pi$ and let $x,y \in C$.  If $y$ is not the winner in all medians of the induced election on $S\cup \{x,y\}$  for all  $S\subseteq I_M(y,x)$, then $(x,y)$ is a  constraint satisfied by $\pi$.  
 \end{lemma}

 \begin{algorithm}[ht]
\caption{LowInf-Pairs}
\label{alg:low-inf}
\footnotesize              \linespread{1.0}\selectfont  
\DontPrintSemicolon 
\KwData{set of candidates $C$,  matrix $[\delta_{uv}]_{u,v\in C}$, constraints $W \subseteq C^2$, interference threshold $r$}
\KwResult{a set of constraints $M \subseteq C^2$ containing $W$}
$M \gets W$; \, $num \gets -1$\; 
\While{$num <|M|$} 
{$num \gets |M|$\; \For{$(x,y)\in C^2\setminus M$ such that $y\neq x$ and $|I_M(x,y)|\leq r$}
{\If{$y$ is not the winner in every median $\pi$ of the election restricted to $\{x,y\}\cup A$ for all $A\subseteq I_M(x,y)$}{ append $(x,y)$ to $M$ \\  
$M \gets \overline{M}$\;
}}
}
\KwRet{$M$}
\end{algorithm}

 If the interference size $|  I_M(y,x)|$ is small (the exact bound depends on practical purposes), we can simply solve all the induced elections on $  S\cup\{x,y\}$ for all subsets $S\subseteq  I_M(y,x)$ to check the condition in Lemma~\ref{l:low-interference}. 
  We observe that  the subproblems are easier for $\alpha$MOT simply because the right hand side of the condition \eqref{t:alpha-beta-2MOT-3.0} has fewer terms. Hence, we can also add potentially much more extra constraints by applying directly  $\alpha$MOT to every subproblem rather than just reusing already found constraints for the original problem.

 \section{Preprocessing II: M-XCC partition} 
 \label{s:preprocessing-II}
 
 The main objective of the preprocessing II step  is to compute the finest $M$-Condorcet partition of the set of candidates (see Definition~\ref{def:M-XCC-partition}) which refines the Finest XCC partition by exploiting a given set of pairwise constraints $M\subseteq C^2$. The existence and uniqueness properties of such a partition are discussed in 
Section~\ref{s:M-XCC-prorperties}. Based on the algorithm for the finest XCC partition in \cite{escobedo}, we introduce  Algorithm~\ref{alg:M-XCC} to compute our finest $M$-Condorcet partition and prove its correctness in Section~\ref{s:algo-M-XCC}. 
  
\subsection{Finest $M$-Condorcet partitions} 
\label{s:M-XCC}
\par 
 We observe that $\alpha$MOT and XCC  do not strictly imply each other. However, simulation results show that XCC adds  only a very small number of extra constraints to $\alpha$MOT and this only happens in very rare cases.  Conversely, constraints obtained by $\alpha$MOT almost always imply the finest XCC partition. To outperform XCC, we introduce below in 
 Definition~\ref{def:M-XCC-partition} the notion of finest $M$-Condorcet partition together with Theorem~\ref{t:XCC-abMOT} explaining how to obtain an XCC type partition using input constraints provided by   $\alpha$MOT and $\alpha$MOTe. When $M$ is the set of constraints obtained by $\alpha$MOT or $\alpha$MOTe, the unique finest $M$-Condorcet partition contains  more blocks and the maximum block size is  smaller comparing to  the  XCC partition.
  
\par 

\begin{definition}  
\label{def:M-XCC-partition} 
Let $(C,V)$ be an election and let $M \subseteq C^2$.  
A partition $C=\bigcup_{i=1}^k X_i$, also denoted $\{X_1,\cdots, X_k\}$,    is called an $M$-Condorcet  partition ($M$-XCC partition) if  for all $x_i\in X_i$ and $x_j \in X_j$ with $i<j$, we have $(x_i,x_j)\in M$ or else $\delta_{x_ix_j}>0$ and $(x_j,x_i)\notin M$. Such a partition with the maximal number of blocks is called a  finest $M$-Condorcet partition.  
\end{definition}

For  $M=\varnothing$ in Definition~\ref{def:M-XCC-partition}, we recover the finest-Condorcet partitions introduced in \cite{escobedo}.   
 
\begin{example}
    \label{ex:toy-example-partition} For the election $(C,V)$ described in Example~\ref{ex:toy-example-1}, we give in Table~\ref{table:example-paritions} below the finest $M$-Condorcet partition where $M$ is the set of constraints obtained by one of the exact methods MOT, $\alpha$MOT, or $\alpha$MOTe. 
\end{example}

  \begin{table}[H]
  \small 
    \centering
       \caption{\small Finest $M$-Condorcet partitions of the election given in Example~\ref{ex:toy-example-1} where the set of constraints $M$ is obtained by the methods described in the first column.}
     \label{table:example-paritions} 
    \begin{tabular}{l c}
      \toprule   
 {$M$}     & Finest $M$-XCC partition   \\ 
\midrule  
$M=\varnothing$ & $\{1,2,3,4,5,7,8\} > \{6\}$ \\ 
MOT & $\{1,2,3,4,5,7,8\} > \{6\}$ \\ 
$\alpha$MOT &  $\{2,3,5,8\}>\{1,7\} >\{4\}> \{6\}$ \\
$\alpha$MOTe \,\, &   $\{2\}>\{5\}>\{8\}>\{3\}>\{1\} > \{7\}>\{4\}> \{6\}$ \\
\bottomrule
    \end{tabular} 
      \vspace{-0.2cm}
  \end{table}

  \subsection{Properties and uniqueness}
 \label{s:M-XCC-prorperties}
The following result generalizes the XCC property discovered by Truchon \cite{truchon-XCC} and thus affirms the usefulness of our $M$-XCC partitions.   In essence, it says that if a median satisfies some set of pairwise constraints  $M\subseteq C^2$ then the sane median must obey  every $M$-XCC partition. 

\begin{theorem}
\label{t:XCC-abMOT}
Let $\pi$ be a median of a constrained election $(C,V,W)$. 
Let $C=\bigcup_{i=1}^k X_i$ be an $M_\pi$-XCC partition, where $M_\pi$ is a set of constraints satisfied by $\pi$. Then $x_i>^\pi x_j$ whenever $x_i\in X_i$ and $x_j \in X_j$ with $i<j$. 
\end{theorem}

\begin{example}
    \label{ex:toy-example-partition-1} By Theorem~\ref{t:XCC-abMOT}, we can infer directly from the last row of Table~\ref{table:example-paritions} that $\pi \colon 2>5>8>3>1>7>4>6$ is a median of the election $(C,V)$ described in Example~\ref{ex:toy-example-1}. Moreover, the partition $\{2,3,5,8\}>\{1,7\} >\{4\}> \{6\}$ must be satisfied by all medians of the same election. 
 \end{example}

Our next result proves the existence and uniqueness of finest $M$-Condorcet partitions under the anti-symmetry  assumption on the set of constraints $M$, for example when $M$ contains only valid constraints satisfied by a median $\pi$.

\begin{theorem}
\label{t:unique-M-XCC}
For every anti-symmetric set $M\subseteq C^2$ of  constraints, the finest $M$-XCC partition exists and is unique. 
\end{theorem}

\subsection{Algorithm for the finest $M$-XCC partition} 
\label{s:algo-M-XCC}

We propose below a quadratic time complexity algorithm  to compute the finest $M$-Condorcet partition (Algorithm~\ref{alg:M-XCC}). When  we set $M=\varnothing$ as an input of Algorithm~\ref{alg:M-XCC}, we recover the Finest-Condorcet Partitioning (XCC) algorithm described in \cite[Algorithm 1]{escobedo}. 
For each $x \in C$,  \cite{escobedo} defines an index  $i_x = |\{y \in C \colon \delta_{xy}>0\}|$ which is  the number  of candidates $y$  to which $x$ is pairwise preferred by the majority of votes. 
The Finest-Condorcet Partitioning algorithm begins by sorting all the candidates $x$ with decreasing index $i_x$ (ties are handled arbitrarily). By this way, they obtain an initial partition $\textbf{X}^{0} = \{X_1^{0},X_2^{0},\dots, X_w^{0}\}$ and a  sequence $I_1>I_2>\dots>I_w$ with $i_x=I_k$ for all $x \in X_k^0$. These  blocks $X_1^{0},X_2^{0},\dots, X_w^{0}$ then enter a successive merging and validating process until an XCC partition is reached. 
 Our Algorithm~\ref{alg:M-XCC} is essentially the same except for the value $i_B$ being replaced by the value $\gamma_B$ that is, for every $x\in B$, the number of candidates $y$ such that $(x,y) \in M$ \textbf{or else} $\delta_{xy}>0$ and $(y,x)\notin M$. It is crucial, notably  for the uniqueness of the finest $M$-Condorcet partition when $M=M_\pi$ contains only constraints satisfied by some median $\pi$,  that the consideration for the constraints in $M$ prevail the majority pairwise preference.  
The technical core of Algorithm~\ref{alg:M-XCC} and its correctness (Theorem~\ref{t:algo-finest-correct}) rely on the  following lemma.

\begin{lemma}
\label{t:index-XCC}
Let $M\subseteq C^2$ be an anti-symmetric and anti-reflexive set of  constraints. 
 For every $x\in C$, let $\gamma_x= |\Gamma_x|$ where 
 $\Gamma_x = \{y \in C  \colon (x,y) \in M   \text{{ or else} } \delta_{xy}>0  \text{ and } (y,x)\notin M \}$. Let $\{X_1,\dots, X_k\}$ be an $M$-XCC partition. Let  $u\in X_i$ and $v\in X_j$ with $i<j$. Then $\gamma_u> \gamma_v$. In particular, if $\gamma_x=\gamma_y$ then $x,y\in X_i$ for some $i$. 
\end{lemma}
 
   \begin{algorithm}[ht]
\caption{Finest-M-Condorcet Partitioning (Finest M-XCC)}
\label{alg:M-XCC}
\footnotesize            
    \linespread{1.0}\selectfont 
\DontPrintSemicolon 
\KwData{candidates $C$, matrix $[\delta_{xy}]$, constraints $M\subseteq C^2$} 
\KwResult{Finest M-XCC partition $\textbf{X}^{f,M}$ and 
$M$-initial partition $\textbf{X}^{0,M}$} 
$\Gamma_x \gets \{y \in C  \colon (x,y) \in M  \text{ \textbf{or else} } \delta_{xy}>0  \text{ and } (y,x)\notin M\}$\;  
$\gamma_x \gets  |\Gamma_x|$  \,  $\forall x \in C$\; 
Construct the partition $\textbf{X}^{0,M}$ whose $i$-th block $X_i^{0,M} \neq \varnothing$ consists of candidates with the $i$-th highest $\gamma$-value\; 
$\widetilde{\textbf{X}} \gets \{X_1^{0,M},X_2^{0,M},\dots, X_w^{0,M}\} = \textbf{X}^{0,M}$; \, $k \gets 1$\; 
\While{$k < |\textbf{X}^{0,M}|$}{
$V  \gets \{ y \in C \setminus    \bigcup_{i=1}^k \widetilde{X}_i \colon \exists x \in \widetilde{X}_k, \, y \notin \Gamma_x\} $\;

\Comment*[r]{elements causing  $M$-XCC violation to $\widetilde{X}_k$} 
\If {$R = \varnothing$}
{$k\gets k+1$  \Comment*[r]{no  $M$-XCC violation   to $\widetilde{X}_k$ } 
} 
\Else{\While{ $V \neq \varnothing$ \Comment*[r]{merge blocks causing $M$-XCC violation} }
{$l\gets \max\{j \colon R \cap \widetilde{X}_j \neq \varnothing \}$\; 
$\widetilde{X}_k \gets \bigcup_{i=k}^l \widetilde{X}_i $\;
$V  \gets \{ y \in C\setminus    \bigcup_{i=1}^k \widetilde{X}_i\colon \exists x \in \widetilde{X}_k, \, y \notin \Gamma_x\} $\; 
}
$k \gets l+1$\; 
}
}
${\textbf{X}}^{f,M}$ $\gets$ $\widetilde{\textbf{X}}$\; 
\KwRet{$\textbf{X}^{f,M}, \textbf{X}^{0,M}$} 
\end{algorithm}

\begin{theorem}
\label{t:algo-finest-correct}
    For  every anti-symmetric   anti-reflexive input  set    $M\subseteq C^2$ of  constraints,  Algorithm~\ref{alg:M-XCC} outputs the unique finest $M$-XCC partition.   
\end{theorem}

\begin{proof}
By the we deal with a block $\widetilde{X}_k$ in Algorithm~\ref{alg:M-XCC} in the outer loop while, we first compute the set $V$ of all $y\in \widetilde{X}_p$ with $p>k$ that causes an $M$-XCC violation to some $x \in \Gamma_x$, that is $y \notin \Gamma_x$. By induction, there is no $M$-XCC violation among the first blocks up to $\widetilde{X}_k$. 
We then merge $\widetilde{X}_k$ with all consecutive blocks up to the last block containing some elements in violation to $\widetilde{X}_k$. This ensures that there is no more violation between the old $\widetilde{X}_k$ and the remaining blocks. However, elements in the  blocks that were merged to $\widetilde{X}_k$ may be in violation with the remaining blocks. For this, the inner while loop updates $V$ and repeats the merging process until no more violation is found. The output $\textbf{X}^{f,M}$ is thus an $M$-XCC partition. 
\par 
Lemma~\ref{t:index-XCC} tells us essentially that every block in an $M$-XCC partition must be the union of consecutive blocks in $\textbf{X}^{0,M}$. Hence, to obtain an finest $M$-XCC partition, we must merge consecutive blocks in $\textbf{X}^{0,M}$ such that there is no more $M$-XCC violation and that the number of resulting blocks is maximized. As Algorithm~\ref{alg:M-XCC} performs only the  necessary merging at each step, the output $\textbf{X}^{f,M}$ must be the finest $M$-XCC partition.  
\end{proof}

As for the original Finest-Condorcet Partitioning algorithm of Akbari and Escobedo \cite{escobedo},   we show that our $M$-Finest-Condorcet Partitioning algorithm also enjoys the time complexity $O(n^2)$.

\begin{theorem}
Algorithm~\ref{alg:M-XCC} has time complexity $O(n^2)$ for every input constraints $M\subseteq C^2$ and every input NP matrix $[\delta_{xy}]_{x,y\in C}$ where $n=|C|$.  
\end{theorem}

\section{Solving: augmented IP formulation} 
\label{s:ip-formulation}
 We now describe the model to incorporate every given set of constraints into the integer programming (IP) formulation of the Kemeny aggregation problem. 
 Every ranking $\pi$  of elements of a set $C$ is uniquely determined by a binary vector $[c_{xy}]_{x,y\in C}$ where $c_{xx}=0$ for all $x\in C$, $c_{xy}=1$ if $x$ is ranked before $y$ in $\pi$, and $c_{yx}=1$ if $y$ is ranked before $x$ in $\pi$. It is clear that such a binary vector satisfies the probability constraints $c_{xy}+c_{yx}=1$ for all distinct $x,y \in C$. By the transitivity of complete orderings, the vector $[c_{xy}]_{x,y\in C}$ also satisfies the transitivity constraints $c_{xy} + c_{yz} + c_{zx} \geq 1$ for all distinct elements $x,y,z\in C$.  This setup constitues  the standard IP formulation of the Kemeny aggregation problem as in \cite{ailon}.
 Suppose that $M\subseteq C^2$ is a set of valid constraints satisfied by some median. Then we can incorporate  the constraints $c_{xy} =1$ whenever $(x,y) \in M$. We obtain the augmented IP formulation to solve for medians satisfying $M$: 
\begin{align}
\label{e:IP}
  \min \quad & \sum_{\{x,y\}  \in C}  c_{xy} a_{yx} + c_{yx} a_{xy} \\
\mathrm{s.t.}\quad  &  c_{xy} \in \{0,1\}, \, c_{xy} + c_{yx} = 1,   \quad  \forall  \{x,y\}  \in C \nonumber  \\ 
& c_{xy} + c_{yz} + c_{zx} \geq 1, \quad \, \forall  \{x,y,z\}  \in C \nonumber 
\\
& c_{xy}  = 1,  \quad  \forall  (x,y) \in M \nonumber  
\end{align}
To obtain a single optimal solution from an empty initial set of constraints $M=\varnothing$, the running time of a generic commercial solver becomes quickly unreasonable and highly unstable as soon as the size $|C|\geq 100$.  
The non-trivial extra cuts $c_{xy}  = 1,  (x,y) \in M$, accelerate tremendously the Branch and Cut algorithm of the IP solvers especially for hard instances of the Kemeny aggregation problem. If we relax the constraints $c_{xy} \in \{0,1\}$  by the corresponding  continuous constraints $0\leq c_{xy} \leq 1$ for all distinct $x,y \in C$, then we obtain the LP relaxations of the above IP formulation \eqref{e:IP}.

\section{Experimental framework}
\label{s:simulation}

To evaluate the computational efficiency and mathematical tightness of the proposed framework, we conduct extensive numerical experiments on both real-world voting profiles and hard synthetic instances. 

\subsection{Experimental setup and data characterization}

The empirical validation relies on two distinct data benchmarks. 

 \medskip 
 \noindent \textbf{Synthetic Mallows Instances.} To analyze scalability under controlled combinatorial complexity, we generate synthetic voting profiles according to the Mallows model \cite{mallows} that we recall below using the Repeated Insertion Method (RIM) \cite{mallows-RIM, effective-mallows}. 
  The Mallows distribution with dispersion factor $\theta \in (0,1]$ is defined by 
$ \mathrm{Prob}(\pi) = \frac{\theta^{d_{KT}(\pi, \pi_0)}}{Z_\theta}$ 
where $\pi_0$ is a given central ranking, $d_{KT}$ is the Kendall-Tau distance, and $Z_\theta$ is the constant normalization factor. When $\theta \to 0$, the probability mass concentrates sharply around $\pi_0$, yielding computationally trivial instances. Conversely, as $\theta \to 1$, the distribution approaches a uniform one, increasing the combinatorial hardness of the problem. 
In the Mallows model, each voter has a probability 
$ p = \frac{1}{1+\theta}$  of correctly ordering any given pair of alternatives (see Table~\ref{table:p-theta} for some values).  
\begin{table}[H]
\centering
\caption{Some values of  $p$ in function of $\theta$}
\label{table:p-theta} 
\begin{tabular}{lcccccc}
\hline\hline
$\theta$ & 0.75   & 0.8    & 0.85   & 0.9    & 0.95   & 0.98   \\ \hline
$p$      & 0.5714 & 0.5556 & 0.5405 & 0.5263 & 0.5128 & 0.5051 \\ \hline\hline
\end{tabular}
 \vspace{-0.3cm}
\end{table} 

\medskip 
\noindent \textbf{Real-world instances.} We test instances consisting of strictly-ordered complete (soc) data from the Preference Library PrefLib \cite{prelib}. For each instance, the dispersion parameter $\theta$ is estimated via the  Maximum Likelihood Estimation (MLE), as formalized in Lemma~\ref{l:theta-estimation}, where every median can be treated as the central reference ranking   in the Mallows model.  
 \medskip 
\noindent  \textbf{Computing the estimation of the dispersion parameter $\theta$.} 
We describe the method to compute the estimated value of  the dispersion parameter according to the MLE in the Mallows model. Let $V=\{\pi_1, \dots, \pi_m\}$ be a voting profile of $m$ rankings over $n\geq 2$ candidates $C$. Let  $
d^*_V= \displaystyle \min_{\sigma\in S(C)} d_{KT}(\sigma, V)$  be the Kemeny score of an arbitrary median $\pi^*$. 
 The MLE value $\theta_{MLE}(V, \pi^*)$ of  $\theta$ in the Mallows model  is the value which maximizes  the realization probability of  $V$. In other words, 
$
\theta_{MLE}(V,\pi^*) \in \mathop{\mathrm{arg\,max}}\limits_{\theta \in (0,1]}  P(X_1=\pi_1, \dots, X_m=\pi_m\vert \pi^*, \theta)$ 
where the random variables $X_1,\dots, X_m$ are  drawn independently from a Mallows distribution with  parameter $\theta$ and with the central ranking $\pi^*$.  
 \begin{lemma}
     \label{l:theta-estimation} 
The value   
     $
     \theta_{MLE}(V,\pi^*)
     $ is independent of the choice of the median   $\pi^*$ and is  denoted by $\theta_{MLE}(V)$. Moreover,  $\theta_{MLE}(V)$ 
     is the unique solution  in $[0, 1)$ of the  strictly  increasing function $H(\theta)$  where   
     $$
     H(\theta)= \sum_{k=1}^n \frac{k}{1-\theta^k} + \frac{n\theta}{1-\theta}-   \frac{n(n+1)}{2} - \frac{d_{V}^*}{m}. 
     $$
 \end{lemma}

\subsection{Evaluated algorithmic configurations}
\label{s:notation-algo} 
For clarity and notation consistency, we define: 
\begin{itemize}
   \item \textbf{MOT:} The baseline constraint set  $M_{\mathrm{MOT}}$  solved by the MOT method.
    \item \textbf{$\pmb{\alpha}${MOT}:} The deterministic constraint set $M_{\mathrm{\alpha MOT}}$ discovered via the exact $\alpha\text{MOT}(C, [\delta_{xy}], \varnothing)$ parallel Algorithm \ref{alg:parallel-abMOT}.
    \item \textbf{$\pmb{\alpha}$MOTe:} The extended deterministic constraint set $M_{\alpha\text{MOTe}}$ derived from the exact $\alpha\text{MOTe}(C, [\delta_{xy}], \varnothing)$ workflow (Algorithm~\ref{alg:abMOTe}). 
    \item $\pmb \alpha$\textbf{MOTe-Inf:} The extended deterministic  constraint set $M_{\alpha \text{MOTe-Inf}}$ built upon $M_{\alpha\text{MOTe}}$ and discovered by LowInf-Pairs (Algorithm~\ref{alg:low-inf}). 
\end{itemize}

\par 
\noindent We also evaluate five distinct finest $M$-Condorcet partitions ($M$-XCC) computed by Algorithm~\ref{alg:M-XCC}: 
\begin{itemize} 
    \item \textbf{XCC:} The baseline finest XCC partition  ($M = \varnothing$).
    \item \textbf{$\pmb{\alpha}$XCC:} The finest $M$-XCC partition restricted by the exact primary active constraint set ($M = M_{\mathrm{\alpha MOT}}$).
    \item \textbf{$\pmb{\alpha}$XCCe:} The finest $M$-XCC partition constrained by the complete admissible extension set ($M = M_{\mathrm{\alpha MOTe}}$). 
      \item \textbf{$\pmb{\alpha}$XCCe-Inf:} The extended finest $M$-XCC partition constrained by the complete admissible extension set ($M = M_{\mathrm{\alpha MOTe-Inf}}$).  
\end{itemize}

\subsection{Evaluation protocol against the baseline}
\label{ss:evaluation-protocol}

We enforce the following strict baseline evaluation protocol to ensure a rigorous and transparent benchmark. All simulations were executed using Python 3.9.6, Apple clang version 21.0.0, and Gurobi 12.0.3 on an Apple M2 MacBook Air (8-core CPU) with 16 GB of unified memory, 256 GB of SSD storage, and macOS Tahoe 26.5.2. 

\medskip
\noindent\textbf{Our framework.}  As in Figure~\ref{fig:pipeline}, we first extract a set of deterministic pairwise preference constraints, denoted as $M$, by executing the parallel Algorithms~\ref{alg:parallel-abMOT}-\ref{alg:abMOTe}  for $\alpha${MOT} and $\alpha${MOTe}  (\textit{Preprocessing I} in Sections~\ref{s:preprocessing-I}-\ref{s:aMOTe-G1-G2}). We then feed $M$ into Algorithm~\ref{alg:M-XCC} to obtain the $M\text{-XCC}$ partition (\textit{Preprocessing II} in Section~\ref{s:preprocessing-II}). Finally, for each block within this partition,  $M$ is leveraged to prune the constraints of the Integer Programming (IP) model \eqref{e:IP}. These models are implemented and solved by Gurobi  via the Python gurobipy API. The internal subproblem optimization step in Algorithm~\ref{alg:Iterated-abMOTe}, requiring the identification of a subset $S \subseteq C$ maximizing $|E(S)|$, is also solved natively through this pipeline. To measure raw algorithmic capability, Gurobi is deployed in a strictly ``naked'' configuration, explicitly disabling its generic internal presolve routines, cutting planes, and primal heuristics ($\textit{Presolve}=0, \textit{Cuts}=0, \textit{Heuristics}=0$) under a sequential single-core computing boundary ($\textit{Threads}=1$). 
    \par 
\medskip
\noindent
\textbf{Implementation of $\pmb \alpha$MOTe-Inf and $\pmb \alpha$XCCe-Inf.} To compute $M=M_{\alpha \text{MOTe-Inf}}$ (see Section~\ref{s:notation-algo}) and the associated $M$-XCC partition denoted $\alpha$XCCe-Inf, we apply Algorithm~\ref{alg:low-inf} with $W=M_{\alpha \text{MOTe}}$ as the set of input constraints and with $r\in \{8,10\}$. Each subproblem of verifying a pair $(x,y)$ with low interference $| I_W(x,y)| \leq r$ is solved sequentially by Gurobi with its default configuration on the corresponding IP model \eqref{e:IP} pruned by $W$. 
    \par 
\medskip
\noindent\textbf{The baseline.} We first compute the XCC partition then construct the standard IP model \eqref{e:IP} for each isolated block without any a priori structural pruning, and invoke the commercial solver Gurobi via gurobipy. In this regime, Gurobi operates under its accelerated default configuration with its generic internal presolve routines, cutting planes, and primal heuristics ($\textit{Presolve}=1, \textit{Cuts}=1, \textit{Heuristics}=1$),  and with ($\textit{Threads}=1$) for benchmarking parity.
 \par 
 \medskip
\noindent\textbf{Timeout protocols.} The total solving time by the solver for each instance is bounded  by $T_{\mathrm{timeout}} = 300\,\text{s}$.

\subsection{Performance indicators} 
\label{ss:performance-indicators}

To evaluate the efficiency, scalability and polyhedral tightness of our algorithms, we monitor a comprehensive hierarchical set of indicators.

\medskip

\medskip
\noindent\textbf{Combinatorial space pruning (\% pairs solved).} The density, in the total space of all $n(n-1)/2$ ordered pairs, of the pairs whose relative order is determined  by MOT, $\alpha$MOT, $\alpha$MOTe, or $\alpha$MOTe-Inf. Maximizing this density compresses the search space by several orders of magnitude. 

\medskip
\noindent\textbf{Decomposition metric (max block size).}  Dictating the worst-case complexity for the branch-and-bound solver, this primary structural hardness indicator  measures the size of the largest block in various Condorcet  partitions.

\medskip
\noindent\textbf{Polyhedral tightness ($\sum$ simplex iterations + 1).} The total accumulated number of simplex pivots executed across all subproblems of the partition. Because the baseline routinely achieves integer convergence at the root node, the traditional node count is an insufficient metric. We use a logarithmic scale rendering. Zero-valued instances, where our constraints force natural integrality at the root, are regularized as $\sum \text{iterations} + 1$.

\medskip
\noindent\textbf{Constraint solving time.} The cumulative wall-clock time required to compute the set of pairwise constraints $M$  and  the $M\text{-XCC}$ partition. 
 
\medskip
\noindent\textbf{Model generation time.} The cumulative wall-clock time requited to generate and load the IP model~\eqref{e:IP} for the Python-Gurobi declaration API.

\medskip
\noindent\textbf{Solver optimization time.} The total wall-clock time required by the Gurobi solver to optimize each subproblem within the partition.  
 
\medskip
\noindent\textbf{Speedup factor analytics.} All acceleration ratios are evaluated on an instance-by-instance basis. 
 To handle instances with zero solving time, runtimes are regularized by introducing an empirical floor defined as half of the strictly positive empirical minimal solving time among tested instances.

\section{Experimental results on synthetic data}
\label{s:results-mallows}
\par
\begin{figure}[H]
\centering
 
    \begin{subfigure}{0.46\textwidth}
     \centering
        \includegraphics[width=\linewidth]{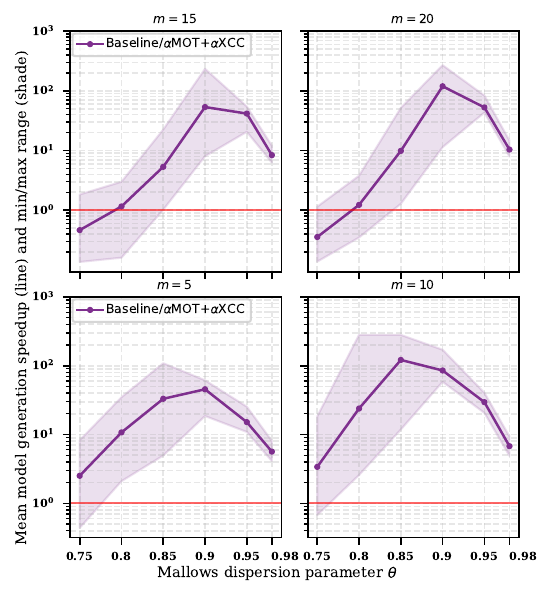}%
\caption{Model  speedup factors achieved by ($\alpha${MOT} + $\alpha$XCC+ naked Gurobi).}
\label{fig:t-model-abmot-speedup}
\end{subfigure}
    \hfill 
  \begin{subfigure}{0.46\textwidth}
     \centering
        \includegraphics[width=\linewidth]{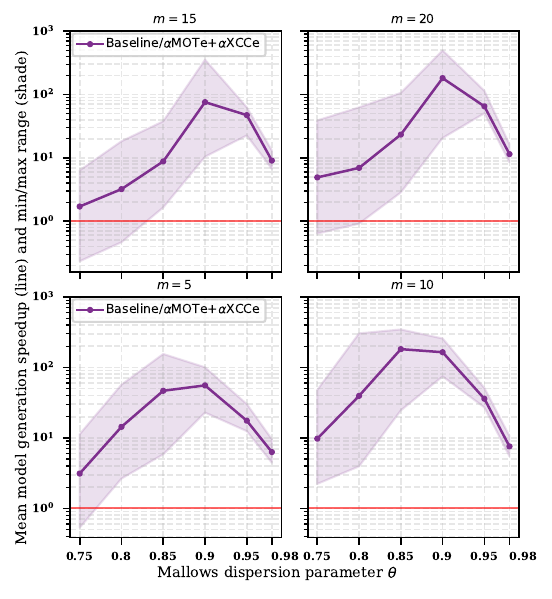}
 
\caption{Model  speedup factors achieved by ($\alpha${MOTe} + $\alpha$XCCe + naked Gurobi).}
\label{fig:t-model-abmote-speedup}
\end{subfigure} 
\caption{Model generation speedup factors achieved by our  frameworks relative to the baseline (XCC + Gurobi presolves, cuts, heuristics).}
\label{fig:t-model-speedup} 
   \end{figure}

\begin{figure}[H]
\centering
 
    \begin{subfigure}{0.46\textwidth}
     \centering
        \includegraphics[width=\linewidth]
{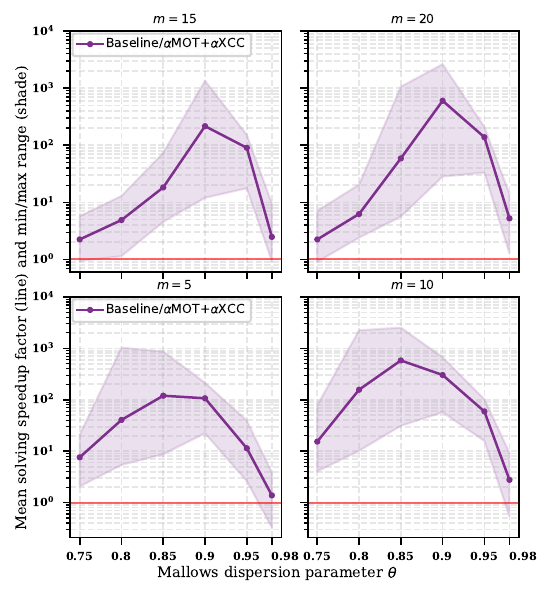}
 
\caption{Solving speedup factors achieved by ($\alpha${MOT} + $\alpha$XCC+ naked Gurobi).}
\label{fig:t-solve-abmot-speedup}
\end{subfigure}
    \hfill 
  \begin{subfigure}{0.46\textwidth}
     \centering
        \includegraphics[width=\linewidth]
{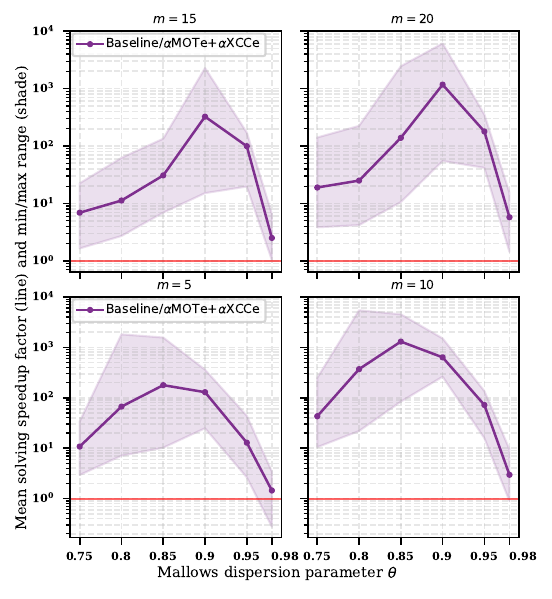}
 
\caption{Solving speedup factors achieved by ($\alpha${MOTe} + $\alpha$XCCe + naked Gurobi) }
\label{fig:t-solve-abmote-speedup}
\end{subfigure}
\caption{Global solving time speedup factors (vertical axis) of our  frameworks relative to the  baseline (XCC + Gurobi presolves, cuts, heuristics)}
\label{fig:t-solve-speedup}
\end{figure}

\begin{figure}[H]
\centering
 
    \begin{subfigure}{0.47\textwidth}
     \centering
        \includegraphics[width=\linewidth]{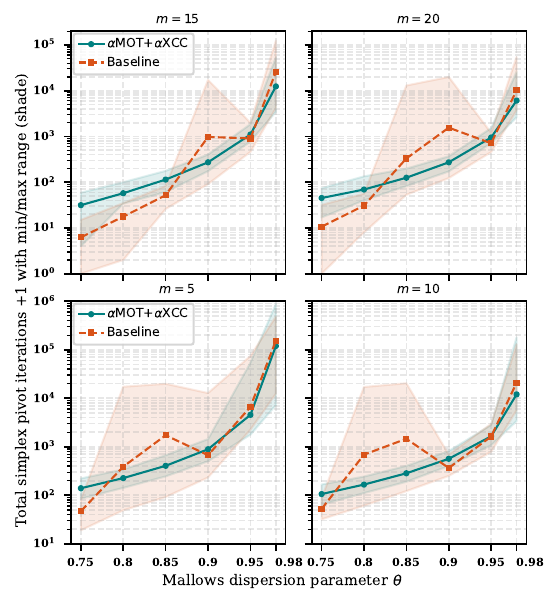}
\caption{($\alpha${MOT} + $\alpha$XCC+ naked Gurobi)}
\label{fig:simplex-iter-abmot}
\end{subfigure}
    \hfill 
 \begin{subfigure}{0.47\textwidth} 
     \centering
        \includegraphics[width=\linewidth]{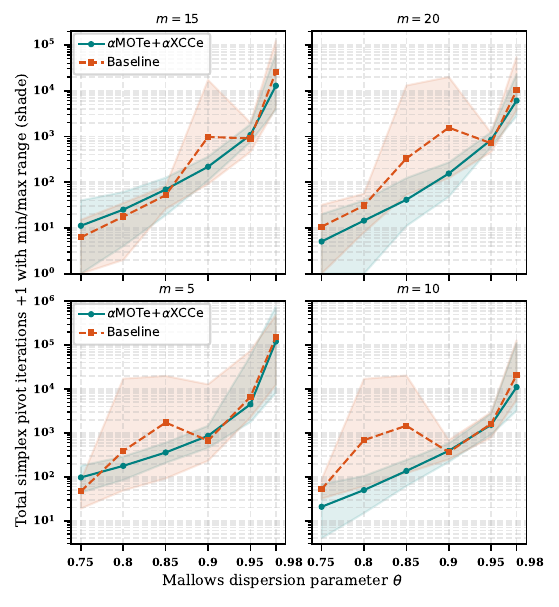}
\caption{ ($\alpha${MOTe} + $\alpha$XCCe+ naked Gurobi)}
\label{fig:simplex-iter-abmote}
\end{subfigure}
\caption{Accumulated number of simplex pivot iterations  of our frameworks   relative to the  baseline (XCC + Gurobi presolves, cuts, heuristics).}
\label{fig:simplex-iter}
\end{figure}

\begin{figure}[H]
\centering
 
    \begin{subfigure}{0.46\textwidth}
     \centering
        \includegraphics[width=\linewidth]{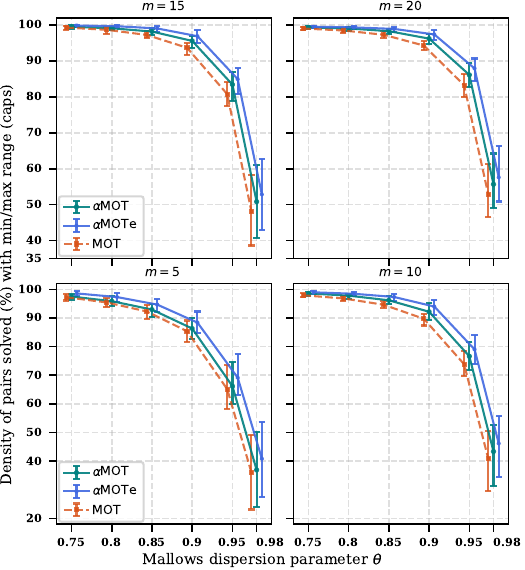}
\caption{Density $(\%)$ of the amount of pairs solved by  MOT, $\alpha$MOT, and $\alpha$MOTe.}
\label{fig:density-pairs-solved}
\end{subfigure}
\hfill 
   \begin{subfigure}{0.46\textwidth}
     \centering
        \includegraphics[width=\linewidth]{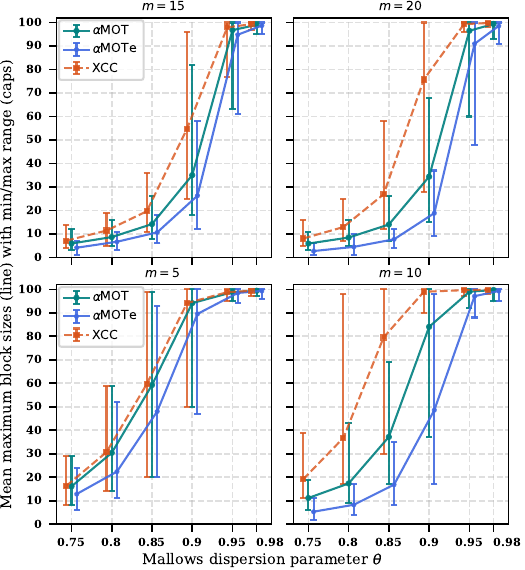}
\caption{Maximum block sizes  in $\alpha$XCC, $\alpha$XCCe, and XCC partitions.}
\label{fig:max-block-size}
\end{subfigure} 
\caption{Density $(\%)$ of the amount of pairs solved (vertical axis) by  MOT, $\alpha$MOT, and $\alpha$MOTe, and maximum block sizes in the Condorcet partitions $\alpha$XCC, $\alpha$XCCe, and XCC.}
\label{fig:ensity-pairs-solved-max-block-size}
\end{figure}

 
We report the speedup factors of the model generation and solving time of our framework (Figures~\ref{fig:t-model-abmot-speedup}-\ref{fig:t-model-abmote-speedup} and Figures~\ref{fig:t-solve-abmot-speedup}-\ref{fig:t-solve-abmote-speedup} ) accompanied by three explicative performance indicators, namely the total simplex pivot iterations (Figures~\ref{fig:simplex-iter-abmot}-\ref{fig:simplex-iter-abmote}), the density of pairwise constraints (Figure~\ref{fig:density-pairs-solved}), and the maximum block sizes in various Condorcet partitions (Figure~\ref{fig:max-block-size}).  
We test instances with $n=100$ candidates,  $m\in \{5,10,15,20\}$ voters, and dispersion parameter  $\theta\in \{0.75, 0.8, 0.85, 0.9, 0.95, 0.98\}$, and generate 100 independent randomized instances per configuration.  Trajectories delineate  means, while the shaded boundaries or the caps illustrate the [Min, Max] intervals across tested instances. The horizontal axis in the graphics represents $\theta$. All the instances tested are solved within 300s by our frameworks and the baseline except for 4 hard instances where $m=5$ and $\theta=0.98$. 
 
\medskip
\noindent\textbf{Model building time speedup analytics.} The highest speedup factors occur when $\theta\in \{0.85, 0.9\}$, ranging 50x-100x on average {(Figure~\ref{fig:t-model-abmot-speedup})} with $\alpha$MOT constraints and $\alpha$XCC partitions. These numbers improve to  60x-200x with $\alpha$MOTe constraints and $\alpha$XCCe partitions (Figure~\ref{fig:t-model-abmote-speedup}).  When $\theta = 0.75$, the XCC partitions become more refined with relative small blocks (see Figure~\ref{fig:max-block-size}) and the gain of our constraints and partitions become less interesting and sometime negative due to overheads.  
When $\theta =0.98$, with almost uniformly random votes in the voting profiles, our framework still boost the model generation running time by a factor 6x-10x.  

\medskip
\noindent\textbf{Solving time speedup analytics.} 
The most important speedup factors also occurs for $\theta\in \{0.85, 0.9\}$   (Figures~\ref{fig:t-solve-abmot-speedup}-\ref{fig:t-solve-abmote-speedup}), ranging 100x-600x and 200x-1000x on average for  ($\alpha$MOT +  $\alpha$XCC + naked Gurobi) and ($\alpha$MOTe + $\alpha$XCCe + naked Gurobi), respectively.  For $\theta=0.75$ and $\theta=0.98$, these frameworks still maintain a competitive  2x-40x and 1.5x-6x  speedup factor.

\medskip 
\noindent \textbf{Polyhedral tightness.}
The framework  ($\alpha$MOT + $\alpha$XCC + naked Gurobi)  significantly reduces the number of pivot iterations in most scenarios, though performance varies by instance hardness (Figures~\ref{fig:simplex-iter-abmot}-\ref{fig:simplex-iter-abmote}).  For hard instances with $\theta \in \{0.95, 0.98\}$, this framework requires only 0.5x to 1.0x the pivot iterations of the baseline. 
When $\theta = 0.9$, it requires slightly more iterations than the baseline for $m \in \{5, 10\}$ but drops to 0.2x–0.3x for $m \in \{15, 20\}$. A similar trend occurs at $\theta = 0.85$, where the framework maintains a 0.2x–0.3x reduction across most settings, except for $m = 15$, which requires 1.5x the baseline iterations. For $\theta = 0.75$, the framework requires up to three times the iterations of the baseline. For $\theta = 0.8$, the framework achieves a 0.3x–0.5x reduction for $m \in \{5, 10\}$ but scales up to 2x–3x the baseline iterations for $m \in \{15, 20\}$. The framework  ($\alpha$MOTe + $\alpha$XCCe + naked Gurobi) exhibits a moderately superior performance.

\medskip 
\noindent \textbf{Pairwise constraint density and Condorcet partition analytics.}
 In general,  $\alpha$MOT and $\alpha$MOTe (Figure~\ref{fig:density-pairs-solved}) yield an extra $1\%\text{--}2\%$ and $5\%\text{--}6\%$ increase in value of the pairwise preference densities, respectively, compared to MOT  when $\theta \in \{0.9, 0.95, 0.98\}$. 

\noindent 
For lower  $\theta \in \{0.75, 0.8, 0.85\}$, these net density margins narrow to $0.5\%\text{--}1\%$ and $2\%\text{--}3\%$. Crucially, within the transitional $\theta \in \{0.85, 0.9\}$ domain, the additional $2\%\text{--}3\%$  density unlocked by $\alpha$MOTe over $\alpha$MOT proves computationally pivotal, enabling $\alpha$XCCe to bisect the average maximum block sizes in the $\alpha$XCC partition (Figure~\ref{fig:max-block-size}). 
\par
\noindent 
An analogous acceleration is observed for both the model generation time and the Gurobi solving time when contrasting  ($\alpha$MOTe + $\alpha$XCCe + naked Gurobi) to the ($\alpha$MOT + $\alpha$XCC + naked Gurobi) framework. For  $\theta = 0.9$ and an even number of voters, the baseline XCC partitions yield a significant largest block averaging $80\text{--}100$ alternatives, downsizing to  $40\text{--}80$  and $20\text{--}50$  alternatives by $\alpha$XCC and $\alpha$XCCe, respectively.

\medskip 
\noindent \textbf{Constraint and partition runtimes.} All runtimes exhibit high predictability 
 with tight standard deviations. For $\theta \in \{0.85, 0.9, 0.95, 0.98\}$, the runtime of  Algorithm~\ref{alg:parallel-abMOT} for $\alpha$MOT remains virtually identical to that of the similar parallel {MOT} formulation (about $0.02-0.03\,\text{s}$). However, for $\theta \in \{0.75, 0.8\}$,  $\alpha$MOT  cuts the baseline {MOT} runtime in half despite solving a larger set of  constraints. The runtime of $\alpha$MOTe requires an additional $50\%\text{--}80\%$ of the core $\alpha$MOT runtime on average. The runtimes of Algorithm~\ref{alg:M-XCC} for all type of Condorcet partitions are similar and scale downward as $\theta$ increases or as more input pairwise constraints are provided, decreasing from approximately $0.001\,\text{s}$ at $\theta = 0.75$ to $0.0002\,\text{s}$ at $\theta = 0.98$.

\section{Experimental results for LowInf-Pairs and an optimized implementation of parallel $\alpha$MOT on large synthetic data} 
   \label{s:results-lowinf} 
   
We evaluate our pairwise constraint discovery engines and partitioning methods including the LowInf-Pairs consideration for large instances with $n=500$ candidates, $m=10$ voters, and interference threshold $r=8$. The results are summarized in Table~\ref{tab:table-n=500-with-LowInf}.   
We observe that running the parallel  Algorithm~\ref{alg:parallel-abMOT} from scratch to compute the set of $\alpha$MOT constraints is not efficient for large $n=500$, notably at the early stage of the execution, which results in a 8x-12x slow down in runtime on average comparing to the similar parallel implementation for MOT. As a solution, we execute the parallel algorithm to obtain the set $M_{\mathrm{MOT}}$ of MOT constraints  then feed $M_{\mathrm{MOT}}$ as input to Algorithm~\ref{alg:parallel-abMOT} to compute the set  $M_{\alpha \mathrm{MOT}}$ of $\alpha$MOT constraints. This optimization reduces the total execution time of $\alpha$MOT to only 2x-3x the runtime of MOT. 
\par 
The cost per new pairwise constraint solved by $ \alpha${MOTe-Inf} may increase significantly and non linearly for hard instances with high dispersion parameter $\theta$  and when we increase the interference threshold to $r=10$ or higher instead of $r=8$. However, the few  additional pairwise constraints unlocked by  $ \alpha${MOTe-Inf} are highly valuable as they cut the maximum block size in the Condorcet partition $\alpha$XCCe by up to 25\%-50\% on average for $\theta \in \{0.85,0.9\}$.  Similarly, for $\theta \in \{0.8, 0.85, 0.9\}$, a mere $0.2\%$-$0.5\%$ increase in the density of pairwise constraints solved by $ \alpha${MOTe-Inf}  with respect to $\alpha$MOT allows the $\alpha$XCCe-Inf partition to bisect the largest block in the $\alpha$XCC partition.  The results for all other performance indicators are coherent with our tests in Section~\ref{s:results-mallows} when $n=100$.

  \begin{table}[H]
\caption{Statistical summary (mean $\pm$ std) of the constraint  discovery and partitioning engines on instances with $n=500$ candidates, $m=10$ voters, dispersion parameter $ \theta \in\{0.75, 0.8, 0.85, 0.9, 0.95\}$, and interference threshold $r=8$.  Each configuration comprises 100 random  instances generated with different seeds. Runtimes are in seconds. $P^{\text{M}}$ ($\%$) is the density of pairs solved by the method M in the space of all $n(n-1)/2$ ordered pairs.  $B^{\text{X}}$ is  the largest block size in the partition X.   
We denote  by  $T_{\text{X}}$ the runtime  to obtain the partition X  
by Algorithm~\ref{alg:M-XCC}.  
$T_{\alpha \text{MOT}}$  is the runtime of Algorithm~\ref{alg:parallel-abMOT} with empty input constraints.  
When we feed the MOT constraints to Algorithm~\ref{alg:parallel-abMOT} to compute the $\alpha$MOT constraints, $T_{\text{M}}^*$ is 
the accumulated runtime to obtain the set of constraints M.}
\label{tab:table-n=500-with-LowInf}
 
\resizebox{\textwidth}{!}{ 
\begin{tabular}{l ccccc}
\toprule
$\theta$ & 0.75   & 0.80   & 0.85   & 0.90   & 0.95     \\
\midrule 
  $P^{\text{MOT}}$  &  99.56 $\pm$ 0.03 & 99.34 $\pm$ 0.03 & 98.89 $\pm$ 0.05 & 97.81 $\pm$ 0.09 & 93.87 $\pm$ 0.16   \\
$P^{\alpha \text{MOT}}$  & 99.69 $\pm$ 0.03 & 99.53 $\pm$ 0.04 & 99.22 $\pm$ 0.05 & 98.32 $\pm$ 0.09 & 94.46 $\pm$ 0.18  \\
$P^{\alpha \text{MOTe}}$ & 99.77 $\pm$ 0.02 & 99.68 $\pm$ 0.02 & 99.46 $\pm$ 0.04 & 98.70 $\pm$ 0.08 & 94.95 $\pm$ 0.19   \\
$P^{\alpha \text{MOTe-Inf}}$ & 99.80 $\pm$ 0.02 & 99.73 $\pm$ 0.02 & 99.57 $\pm$ 0.04 & 98.87 $\pm$ 0.09 & 94.96 $\pm$ 0.19  \\ 
\midrule 
$T_{\text{MOT}}$   & 0.1101 $\pm$ 0.0441 & 0.1318 $\pm$ 0.0359 & 0.1456 $\pm$ 0.0470 & 0.1632 $\pm$ 0.0535 & 0.1613 $\pm$ 0.0502   \\
$T_{\alpha \text{MOT}}^*$ & 0.2468 $\pm$ 0.0650 & 0.2817 $\pm$ 0.0633 & 0.3312 $\pm$ 0.0801 & 0.4325 $\pm$ 0.1031 & 0.3905 $\pm$ 0.0699   \\
$T_{\alpha \text{MOTe}}^*$ & 0.3383 $\pm$ 0.0702 & 0.3827 $\pm$ 0.0673 & 0.4477 $\pm$ 0.0920 & 0.5971 $\pm$ 0.1131 & 0.5781 $\pm$ 0.0844   \\
$T_{\alpha \text{MOTe-Inf}}^*$ & 0.4951 $\pm$ 0.1013 & 0.8590 $\pm$ 0.1937 & 2.2981 $\pm$ 0.4073 & 7.0094 $\pm$ 0.7244 & 2.1223 $\pm$ 0.4029 \\
\midrule 
$T_{\alpha \text{MOT}}$  & 1.1838 $\pm$ 0.0506 & 1.1693 $\pm$ 0.0588 & 1.2228 $\pm$ 0.0618 & 1.3139 $\pm$ 0.0817 & 1.3212 $\pm$ 0.0895  \\
\midrule 
$B^{\text{XCC}}$ & 27.04 $\pm$ 6.08 & 58.10 $\pm$ 17.38 & 205.61 $\pm$ 69.95 & 484.22 $\pm$ 47.25 & 499.66 $\pm$ 0.65  \\
$B^{\alpha \text{XCC}}$ & 16.45 $\pm$ 4.81 & 28.98 $\pm$ 9.30 & 58.28 $\pm$ 16.83 & 271.59 $\pm$ 97.66 & 498.97 $\pm$ 1.62  \\
$B^{\alpha \text{XCCe}}$ & 12.44 $\pm$ 2.53 & 19.62 $\pm$ 4.52 & 38.57 $\pm$ 10.09 & 155.44 $\pm$ 56.07 & 498.27 $\pm$ 2.15   \\
$B^{\alpha \text{XCCe-Inf}}$& 11.27 $\pm$ 1.95 & 16.56 $\pm$ 3.56 & 29.80 $\pm$ 8.77 & 100.28 $\pm$ 35.04 & 495.74 $\pm$ 19.55   \\
\midrule
$T_{\text{XCC}}$ & 0.0362 $\pm$ 0.0032 & 0.0558 $\pm$ 0.0112 & 0.1404 $\pm$ 0.0484 & 0.1707 $\pm$ 0.0194 & 0.0574 $\pm$ 0.0042 \\
$T_{\alpha \text{XCC}}$  & 0.0310 $\pm$ 0.0022 & 0.0378 $\pm$ 0.0051 & 0.0542 $\pm$ 0.0115 & 0.1463 $\pm$ 0.0541 & 0.0690 $\pm$ 0.0046  \\
$T_{\alpha \text{XCCe}}$  & 0.0285 $\pm$ 0.0012 & 0.0317 $\pm$ 0.0020 & 0.0413 $\pm$ 0.0060 & 0.0968 $\pm$ 0.0303 & 0.0750 $\pm$ 0.0055  \\
$T_{\alpha \text{XCCe-Inf}}$  & 0.0277 $\pm$ 0.0011 & 0.0298 $\pm$ 0.0016 & 0.0356 $\pm$ 0.0041 & 0.0675 $\pm$ 0.0203 & 0.0765 $\pm$ 0.0126  \\
\bottomrule

\end{tabular}
}
\end{table}

\section{Experimental results on real-world data}
   \label{s:results-preflib}

Our frameworks are tested for all strict order complete (soc) data in the PrefLib database. For this test, we run Algorithm~\ref{alg:parallel-abMOT} with empty input constraints to compute the set of $\alpha$MOT constraints as in Section~\ref{s:results-mallows}.  As the baseline encounters a severe bottleneck in both model generating and solving time and exceeds frequently the 300s timeout, we restrict the test for the baseline to instances with maximum block size at most $200$ in the XCC partition. There are exactly  4 instances namely  00050-00000001.soc, 00044-00000018.soc, 00044-00000019.soc, and 00044-00000038.soc where our frameworks exceed the 300s solving time allocation. All these instances have a largest block bigger than 200 in size in the XCC partition.  In general, the results (Table~\ref{table:real-data-baseline}) are coherent with our observations on synthetic data discussed in Section~\ref{s:results-mallows}. We observe that the  estimated value $\theta_{MLE}$ of the dispersion parameter of most of the instances with $n> 80$ concentrates near the range $[0.8, 0.95]$ where our methods prove to be most efficient.

 \section{Discussion and perspectives} 
 \label{s:discussion}
For large $n$ (hundreds or more), our  implementation to compute in parallel the set of MOT, $\alpha$MOT, and $\alpha$MOTe constraints enjoys a speedup factor, with respect to the sequential computation of the same constraints, almost equal to the number of parallel working cores. For lower values of $n$,  overhead of parallel computing weighs more in the runtime and the above speedup factor is smaller. 
For industrial applications, a parallel computation of Low-Inf after $\alpha$MOTe will be highly efficient for two reasons. First, the extra pairwise constraints can help to drastically reduce the size of the maximum block in the resulting $M$-XCC partition (see Table~\ref{tab:table-n=500-with-LowInf}). Second, the highly predictable computing cost can be reduced by a linear factor proportional to the number of parallel workers running Low-Inf. Since Low-Inf only solves subproblems of small size bounded by the interference threshold, this parallel computation of Low-Inf features a  desirable divide-and-conquer exact approach.  
To fully optimize the runtime, we recommend running the parallel implemention of MOT first to obtain the set  $M$  of pairwise constraints MOT, then computing the corresponding $M$-XCC partition, applying the method $\alpha$MOT to each block of this $M$-XCC partition, and finally assembling to achieve the set of $\alpha$MOT constraints. See Table~\ref{tab:table-n=500-with-LowInf} for an example of runtime speedup benefit.  
\par 
We note that $\alpha$MOT and $\alpha$MOTe remain valid in the more general context of  weighted tournament where the weights $a_{xy}$ are only required to satisfy the probability condition $0 \leq a_{xy} \leq 1$ and $a_{xy}+a_{yx}=1$ for all distinct $x,y$.  Our methods can be applied verbatim in this more general context of weighted tournaments. 
 Other significant applications of our results  include quick, reliable, and explainable large-scale approximate solutions of the general constrained Kemeny aggregation problem. 
Another perspective is to extend our methods in a straightforward manner to tackle incomplete
rankings or complete rankings with ties and to compare with other
approximate methods such as Soft Condorcet Optimization \cite{soft-condorcet}.

 \begin{table}[H]
 \caption{Results for the framework ($\alpha${MOTe} + $\alpha$XCCe+ naked Gurobi) and the baseline on PrefLib instances  with at least $20$ alternatives and whose XCC maximum block size is at least 3. The instances are grouped by ranges of the number of alternatives $n=|C|$ then by ranges of the estimated dispersion $\theta_{MLE}$. In each range of $n$,  the average number of alternatives and voters are  $\bar{n}$ and $\bar{m}$.  Size is the
number of instances in each group. Each row reports the  mean values for each group. The superscripts $^\text{e}$ and $^\text{b}$ represent our framework and the baseline respectively. $B^{\text{xcc}}$ and $B^e$ are the maximum block size in the XCC and the $\alpha$XCCe partitions.  
 $P^\text{e}$ ($\%$) is the density of pairs solved by $\alpha$MOTe in the space of all $n(n-1)/2$ ordered pairs.  $T^\text{e}_{\text{pre}}$ is the combined runtime to compute $\alpha$MOTe  constraints and the   $\alpha$XCCe partition.  $T_\text{mod}$, $T_\text{sol}$, and $S$  are  the IP model generation time, the Gurobi solving time, and the total number of simplex iterations. Times are in seconds.}  
\label{table:real-data-baseline}
\centering
\footnotesize
\resizebox{\textwidth}{!}{
\begin{tabular}{llccccccccccc}
\toprule
$\theta_{MLE}$  (mean) & size & $B^{\text{xcc}}$ & $B^e$ & $P^\text{e}$ & $T^\text{e}_{\text{pre}}$ & $T^e_\text{mod}$ & $T^e_\text{sol}$ & $S^\text{e}$ & $T^b_\text{mod}$ & $T^b_\text{sol}$ & $S^b$ \\
 \midrule
 \midrule
\multicolumn{12}{l}{\textbf{3845 instances with $3\leq B^{\text{xcc}} \leq 200$}} \\
\midrule
\multicolumn{12}{l}{\textbf{Interval $n$: 20-80} \quad ($\bar{n}=35$, \, $\bar{m}=34$)  } \\
\midrule
0.0-0.6 ($0.55$) & 88 & $5.8$ & $3.0$ & $98.64$ & $0.0129$ & $0.0024$ & $0.0002$ & $6.4$ & $0.0029$ & $0.0010$ & $2.5$ \\
0.6-0.7 ($0.66$) & 320 & $7.9$ & $3.5$ & $98.55$ & $0.0136$ & $0.0022$ & $0.0002$ & $7.2$ & $0.0060$ & $0.0018$ & $4.6$ \\
0.7-0.8 ($0.73$) & 113 & $21.5$ & $12.4$ & $95.68$ & $0.0219$ & $0.0066$ & $0.0007$ & $48.2$ & $0.1219$ & $0.0426$ & $36.9$ \\
0.8-0.9 ($0.85$) & 59 & $52.1$ & $44.0$ & $79.86$ & $0.0395$ & $0.0469$ & $0.0363$ & $594.5$ & $1.0480$ & $1.1733$ & $3595.1$ \\
0.9-1.0 ($0.91$) & 6 & $71.2$ & $66.5$ & $69.56$ & $0.0429$ & $0.1458$ & $0.0886$ & $1283.7$ & $2.2128$ & $2.1649$ & $4436.7$ \\
\midrule
\multicolumn{12}{l}{\textbf{Interval $n$: 81-120} \quad ($\bar{n}=115$, \,  $\bar{m}=17$)  } \\
\midrule
0.0-0.6 ($0.51$) & 36 & $32.2$ & $29.9$ & $98.11$ & $0.0286$ & $0.0300$ & $0.0033$ & $224.3$ & $0.6756$ & $0.5484$ & $1064.2$ \\
0.6-0.7 ($0.67$) & 79 & $26.7$ & $17.7$ & $98.79$ & $0.0362$ & $0.0207$ & $0.0023$ & $134.5$ & $0.8517$ & $0.6115$ & $521.1$ \\
0.7-0.8 ($0.76$) & 471 & $38.3$ & $18.7$ & $98.71$ & $0.0522$ & $0.0203$ & $0.0021$ & $136.5$ & $0.9444$ & $0.6115$ & $927.7$ \\
0.8-0.9 ($0.83$) & 714 & $71.8$ & $36.8$ & $96.79$ & $0.0653$ & $0.0367$ & $0.0061$ & $314.2$ & $3.2744$ & $2.0171$ & $1331.2$ \\
0.9-1.0 ($0.92$) & 51 & $96.9$ & $90.3$ & $75.95$ & $0.0629$ & $0.2684$ & $0.6051$ & $3201.0$ & $5.7247$ & $4.3814$ & $4595.5$ \\
\midrule
\multicolumn{12}{l}{\textbf{Interval $n$: 121-160} \quad ($\bar{n}=131$,  \, $\bar{m}=18$)  } \\
\midrule
0.0-0.6 ($0.51$) & 17 & $45.0$ & $43.9$ & $97.45$ & $0.0287$ & $0.0431$ & $0.0053$ & $371.2$ & $1.2497$ & $0.7085$ & $876.6$ \\
0.6-0.7 ($0.66$) & 31 & $45.8$ & $31.9$ & $98.05$ & $0.0432$ & $0.0387$ & $0.0046$ & $275.0$ & $1.8139$ & $1.2806$ & $765.4$ \\
0.7-0.8 ($0.77$) & 408 & $35.6$ & $16.1$ & $99.14$ & $0.0626$ & $0.0179$ & $0.0016$ & $109.3$ & $0.7552$ & $0.4548$ & $668.9$ \\
0.8-0.9 ($0.84$) & 511 & $75.2$ & $39.0$ & $97.33$ & $0.0716$ & $0.0387$ & $0.0075$ & $326.8$ & $3.9293$ & $2.4135$ & $960.3$ \\
0.9-1.0 ($0.94$) & 36 & $133.1$ & $126.5$ & $75.04$ & $0.1093$ & $0.6601$ & $1.2171$ & $5314.8$ & $14.6477$ & $12.5387$ & $7728.0$ \\
\midrule
\multicolumn{12}{l}{\textbf{Interval $n$: 161-700} \quad ($\bar{n}=307$, \, $\bar{m}=45$)  } \\
\midrule
0.0-0.6 ($0.51$) & 14 & $63.5$ & $63.5$ & $99.08$ & $0.2605$ & $0.1173$ & $0.0145$ & $849.2$ & $5.9024$ & $3.9789$ & $222.1$ \\
0.6-0.7 ($0.65$) & 21 & $61.1$ & $55.3$ & $99.17$ & $0.3148$ & $0.1146$ & $0.0137$ & $719.4$ & $7.5678$ & $6.0547$ & $1713.3$ \\
0.7-0.8 ($0.76$) & 71 & $80.5$ & $44.7$ & $99.15$ & $0.4553$ & $0.0793$ & $0.0095$ & $637.9$ & $9.4881$ & $7.1786$ & $1913.0$ \\
0.8-0.9 ($0.87$) & 600 & $104.5$ & $58.6$ & $98.71$ & $0.3963$ & $0.1029$ & $0.0181$ & $768.6$ & $15.0843$ & $11.7830$ & $2387.5$ \\
0.9-1.0 ($0.92$) & 199 & $125.5$ & $78.1$ & $97.21$ & $0.3816$ & $0.1930$ & $1.6567$ & $3172.4$ & $21.5844$ & $19.0538$ & $3640.0$ \\
 \midrule
 \midrule
\multicolumn{12}{l}{\textbf{1467 instances with $B^{\text{xcc}} \geq 201$ (excluding 4 instances exceeding the 300s solving time)} }   \\
\midrule
\multicolumn{12}{l}{\textbf{Interval $n$: 200-320}  \quad ($\bar{n}=256$, \, $\bar{m}=13$)} \\
\midrule
0.0-0.8 ($0.77$) & 2 & $262.0$ & $186.5$ & $97.20$ & $0.2592$ & $0.5347$ & $0.0935$ & $2713.0$ & --- & --- & --- \\
0.8-0.9 ($0.88$) & 192 & $226.7$ & $128.5$ & $97.23$ & $0.2791$ & $0.2306$ & $0.0456$ & $1376.3$ & --- & --- & --- \\
0.9-1.0 ($0.92$) & 431 & $232.3$ & $183.5$ & $94.30$ & $0.2903$ & $0.5478$ & $0.6241$ & $2814.3$ & --- & --- & --- \\
\midrule
\multicolumn{12}{l}{\textbf{Interval $n$: 321-1080}  \quad ($\bar{n}=356$, \, $\bar{m}=15$) } \\
\midrule
0.0-0.8 ($0.76$) & 9 & $274.1$ & $173.3$ & $97.85$ & $0.3927$ & $0.4960$ & $0.1004$ & $2583.3$ & --- & --- & --- \\
0.8-0.9 ($0.88$) & 266 & $271.1$ & $135.7$ & $98.01$ & $0.6433$ & $0.3184$ & $0.0770$ & $1852.6$ & --- & --- & --- \\
0.9-1.0 ($0.92$) & 569 & $313.8$ & $225.2$ & $96.37$ & $1.2768$ & $1.1901$ & $1.0738$ & $3773.4$ & --- & --- & --- \\
\bottomrule
\end{tabular}
} 
\end{table}

  \begingroup
\setstretch{1.0}  \footnotesize    
\bibliographystyle{elsarticle-harv}  \bibliography{Kemeny}    
\endgroup

\clearpage 

 \appendix

   In the appendices below, we provide the full proofs of the technical results. Theorems, lemmas, algorithms, and references maintain the exact same numbering as in the main paper for ease of reading. 
      
      \setcounter{theorem}{1}
      
\section{Section 3.1: proof of Remark 1 ($\pmb \alpha$MOT generalizes MOT)}
\label{s:aMOT}
 
We claim that the  iterated version of MOT  can be obtained as a particular case of Theorem~1 if we fix $W=\varnothing$ and  $\alpha=\beta=0.5$ in the  condition~(3.1) in the main paper which results in  the following stronger condition 
\begin{align*}
\label{eq:generalize-mot}
 \delta_{xy}> 0.5\sum_{z\in\,  Z_k(x,y)}  \max\left(0,  \delta_{yz} + \delta_{zx} \right).
 \end{align*} 
To see why this condition is equivalent to the description of MOT in Section~2.1 in the main paper in the context of an election, it suffices to show for the base case $k=-1$ (the general case is similar) that  
$$|E_{yx} \setminus E_{xy}| = 0.5 \sum_{z\in\,  C}   \max\left(0,  \delta_{yz} + \delta_{zx} \right).$$
Indeed, for all distinct elements $u,v,w$, let $n_{uvw}$ denote the number of votes in which $u>v>w$. Similarly, let $n_{uv}$ be the number of votes in which $u>v$. Then it is clear that for all distinct elements $u,v,w$, we have 
$$
 ma_{uv} =n_{uv} = n_{wuv}+ n_{uwv}+ n_{uvw}.  
$$
Hence, it follow that from the definition of $E_{yx}$ and $E_{xy}$ that: 
\begin{align*}
2 |E_{yx} \setminus E_{xy}|  & =  \sum_{z\in C}2  \max(0, n_{yzx} - n_{xzy})  \\
& = \sum_{z\in C} \max(0, 2n_{yzx} - 2n_{xzy}) 
\\& = \sum_{z\in C} \max(0, n_{yzx} + n_{xyz} + n_{yxz}  - n_{xzy}- n_{zxy} - n_{zyx}  \\
& + n_{yzx} +n_{zyx} + n_{zxy} - n_{yxz} - n_{xyz} - n_{xzy}  ) \\
& =     \sum_{z\in C} \max(0, m a_{yz} - ma_{zy}  + m a_{zx} - ma_{xz}  ) \\
& =   \sum_{z\in C} \max(0, \delta_{yz}  + \delta_{zx} ). 
\end{align*}

\section{Section 3.2: complexity of computing $\alpha$MOT constraints and the recursive integer-only conditions to compute it} 
 
 \begin{theorem}
In Algorithm $\alpha \mathrm{MOT}\text{-}\mathrm{SinglePair}$, the time complexity for the verification of a pair of candidates $(x,y)$ is  $O(|Z_k|^2) $. Here,  the set $Z_k=Z_k(x,y)$ is given by the formula (3.2) in the main paper and satisfies $|Z_k|\leq n-2$. 
\end{theorem}

\begin{proof}
Computing $ \min \sum_{z\in Z_k}  \max\left(0, \alpha \delta_{yz} + \beta \delta_{zx} \right)$ over $\alpha, \beta\geq 0$ with    $\alpha+\beta=1$ reduces to the computation of  $\min   F_k(\alpha)$ over $\alpha\in [0,1]$. By the equality~(3.3) in the main paper
\begin{align*}
  \min_{\alpha\in [0,1]}   F_k(\alpha)=\min_{1 \leq i\leq r} F_k(t_i)=\min_{t\in T_k} F_k(t), 
\end{align*}
and by the definition 
\begin{align*} 
F_k(t_i) & =  \sum_{z\in Z_k(x,y)}  \max\left(0, t_i  \delta_{yz} + (1-t_i) \delta_{zx} \right), 
\end{align*}
the time required for this computation is  $O(|Z_k||T_k|)$  which is  $O(|Z_k|^2)$  since $|T_k| \leq |Z_k|+2$. 
\end{proof}

\begin{lemma}(Integer-only equivalent conditions)
Fix $k \in \mathbb{N}^*$. Let $T_k = \{t_1, \dots, t_r\}$ with $\displaystyle t_i = \frac{a_i}{b_i}$, where $a_i, b_i \in \mathbb{N}^*$ and $b_i \neq 0$. Let $\displaystyle f_j = \min_{1 \leq i \leq j} F_k(t_i)$. Then $\displaystyle f_r = \min_{\alpha \in [0,1]} F_k(\alpha)$, and we can write $\displaystyle f_j = \frac{\mathcal{N}_j}{\mathcal{D}_j}$, where $\mathcal{N}_j, \mathcal{D}_j \in \mathbb{N}$ can be computed recursively on $j$ using only integer operations as follows:
\begin{align*} 
N_j & = \sum_{z \in Z_k(x,y)} \max\left(0, a_j \delta_{yz} + (b_j - a_j) \delta_{zx} \right), \\
(\mathcal{N}_{j+1}, \mathcal{D}_{j+1}) & =  
\begin{cases}
(N_{j+1}, b_{j+1}) & \mathrm{if\ } N_{j+1} \mathcal{D}_j < \mathcal{N}_j b_{j+1}, \\ 
(\mathcal{N}_j, \mathcal{D}_j) & \mathrm{otherwise.}   
\end{cases} 
\end{align*} 
\end{lemma}

\begin{proof}
The equality $\displaystyle f_r = \min_{\alpha \in [0,1]} F_k(\alpha)$ follows from the  relation (3.3) established in the main text 
\begin{align*}
  \min_{\alpha\in [0,1]}   F_k(\alpha)=\min_{1 \leq i\leq r} F_k(t_i)=\min_{t\in T_k} F_k(t)= f_r. 
\end{align*}
The correctness of the recursive algorithm to compute $f_1, f_2, \dots, f_r$ results immediately from the fact that $\displaystyle t_j=\frac{a_j}{b_j}  $ and 
\begin{align*}
F_k(t_j) & =  \sum_{z\in Z_k(x,y)}  \max\left(0, t_j  \delta_{yz} + (1-t_j) \delta_{zx} \right) \\
& =  \sum_{z\in Z_k(x,y)}  \max\left(0, \frac{a_j}{b_j}   \delta_{yz} + \left ( 1-\frac{a_j}{b_j} \right  ) \delta_{zx} \right)\\
& = \frac{1}{b_j} \sum_{z \in Z_k(x,y)} \max\left(0, a_j \delta_{yz} + (b_j - a_j) \delta_{zx} \right)  = \frac{N_j}{b_j}.
\end{align*}
The proof is thus complete. 
\end{proof} 

 \section{Section 4.1: correctness of $\alpha$MOT{e}}
 \label{s:aMOTe-G1-G2}

 \begin{theorem}[$\alpha$MOT{e}]
 \label{alg:abmote}
Let $E$ be the set of all ordered pairs realizing the equality in the  condition~(3.1) in the main paper for at least one value of $k$. Then for every  $S \subseteq C$, there exists some median satisfying all the constraints in $E(S)=\{(x,y)\in E\colon x\in S, \, y \notin S\}$. 
\end{theorem}

\begin{proof}
Let $S\subseteq C$. Suppose, on the contrary, that no median satisfies all the constraints in $E(S)$. We can thus choose such a median $\pi$ which violates the smallest number of constraints in $E(S)$. Take $(x,y)\in E(S)$  such that $y>^\pi x$ and the number of candidates between $y$ and $x$ in $\pi$ is minimal. We write $\pi \colon L>y>Z>x>R$  where $L,Z,R$ are ordered sets of candidates.  As the number of candidates between $y$ and $x$ in $\pi$ is minimal,  $(z,y)\notin E(S)$ and $(x,z)\notin E(S)$ for all $z \in Z$. Because $(x,y)\in E(S)$, we have $x\in S$ and $y \notin S$ and thus $(z,x)\notin E(S)$ and $(y,z)\notin E(S)$ for all $z \in Z$. 
\par 
Let  $\sigma_1 \colon L>Z>x>y>R$ and $\sigma_2 \colon L>x>y>Z>R$. Since $(x,y)\in E$, the same proof, \textit{mutatis mutandis}, of Theorem~1 shows that $d_{KT} (\pi, V) \geq \min ( d_{KT}(\sigma_1, V), d_{KT}(\sigma_2, V))$. As $\pi$ is a median, it follows that $\sigma_1$ or $\sigma_2$ is also a median. As $(y,z), (z,y)\notin E(S)$ and $(z,x), (x,z)\notin E(S)$ for all $z \in Z$, we deduce that both $\sigma_1$ and $\sigma_2$ satisfy the same constraints in $E(S)$ as $\pi$ plus one additional constraint $(x,y)\in E(S)$. This contradicts the choice of $\pi$ and the theorem is proved. 
\end{proof}
   
 \section{Section 4.2: correctness of LowInf-Pairs} 
 \label{s:low-inf} 
 
 \begin{lemma} 
\label{l:induced-election}
A ranking $\pi$ is a median of the election $(C,V)$ if and only for every subset $S \subseteq C$ of consecutive elements in $\pi$, the restriction $\pi_S$ is a median of the induced election $(S,V\vert_S)$ on $S$ where $V\vert_S$ is obtained from $V$ by deleting all the candidates in $C\setminus S$ in every vote.
\end{lemma}

\begin{proof}
  Indeed, suppose that $\pi$ is a median of $(C,V)$ but $\pi_S$ is not a median of $(S,V\vert_S)$ for some set $S\subseteq C$ of consecutive elements in $\pi$. Let $\sigma$ be the ranking obtained from $\pi$ by replacing the subranking $\pi_S$ by a median of $(S,V\vert_S)$. Therefore, $\sigma_S$ is a median of $(S,V\vert_S)$. Since $\pi_S$ is not a median, $d_{KT}(\pi_S, V\vert_S) > d_{KT}(\sigma_S, V\vert_S)$. As $S$ contains consecutive elements in the ranking $\pi$, it follows that  
\begin{align*}
d_{KT}(\pi, V) - d_{KT}(\sigma, V) = d_{KT}(\pi_S, V\vert_S) - d_{KT}(\sigma_S, V\vert_S)  > 0.
\end{align*}
This means that $\pi$ is not a median of $(C,V)$ which is a contradiction. The converse implication is trivial by considering $S=C$. 
  \end{proof}
  
 We can now provide he proof of the correctness of the method LowInf-Pairs. 
 
 \begin{lemma} 
 \label{l:low-interference}
 Let $M\subseteq C^2$ be a set of constraints satisfied by some median $\pi$ and let $x,y \in C$.  Suppose that $y$ is not the winner in every median of the induced election on $S\cup \{x,y\}$ for every  $S\subseteq I_M(y,x)$. Then $(x,y)$ is a  constraint satisfied by $\pi$.  
 \end{lemma} 
 
 \begin{proof} 
 Suppose on the contrary that $y$ is ranked before $x$ in $\pi$. Let $S$ be the set of all elements ranked between $y$ and $x$ in $\pi$. Since $M$ contains only valid constraints satisfied by $\pi$, 
every element $z \in S$ then satisfies $(z,y)\notin M$ and $(x,z)\notin M$. It follows that $S\subseteq M$.  For $E= S\cup\{x,y\}$, Lemma~\ref{l:induced-election}  together with the fact that  $\pi$ is a median imply that $\sigma= \pi_{E}$ is a median of the induced election on $E$. As $y$ is the winner of $\pi_E$, we obtain a  contradiction to the hypothesis. The proof is thus complete.  
 \end{proof}

 \section{Section 5.2: properties and uniqueness of M-XCC partitions} 
 \label{s:preprocessing-II}
 
\begin{theorem}
\label{t:XCC-abMOT}
\label{t:XCC-abMOT}
Let $\pi$ be some median of a constrained election $(C,V,W)$.  
Suppose that $C=\bigcup_{i=1}^k X_i$ is an $M_\pi$-XCC partition, where $M_\pi\subset C^2$ is a set of constraints satisfied by $\pi$. Then $x_i>^\pi x_j$ whenever $x_i\in X_i$ and $x_j \in X_j$ with $i<j$. 
\end{theorem}

\begin{proof} We proceed by induction on $k\geq 2$ (there is nothing to prove when $k=1$). For $k=2$,  suppose on the contrary that there exist $x\in X_1$ and  $y \in X_2$ such that  $y >^\pi x$. Let $(x,y)$ be such a pair with the fewest candidates  between them in the ranking $\pi$. By hypotheses, it is then clear that $y$ and $x$ must be adjacent in that order in $\pi$ since every candidate $z$ lying between $x$ and $y$ in $\pi$ will create a pair, either with $x$ or $y$, that has even fewer candidates in between. Moreover, $(x,y)\notin M_\pi$ as $\pi$ satisfies $M_\pi$. Since $C=  X_1\cup X_2$   is an $M_{\pi}$-XCC partition, it follows that  $\delta_{xy}>0$. Swapping $y$ and $x$ in $\pi$ will decrease the Kemeny score of $\pi$ by $\delta_{xy}>0$ and the resulting ranking still satisfies $M_\pi$ and the initial constraints $W$ since $W\subset M_\pi$. 
We thus obtain a contradiction since  $\pi$ is a median of the constrained weighted tournament. Hence, the theorem holds when $k=2$. Now suppose that the result is true for some $k \geq 2$. Let $C=\bigcup_{i=1}^{k+1} X_i$ be an $M_\pi$-XCC partition. Let $X=\bigcup_{i=1}^{k} X_i$ then  $C=X\cup X_{k+1}$ is an $M_\pi$-XCC partition. Hence, $x >^\pi y$ for all $x\in X$ and $y \in X_{k+1}$. In particular, the candidates in $X$ are adjacent in $\pi$ and it follows that the restriction of $\pi$ on $X$ induces a median $\pi_X$ of the  resulting election on $X$ where we simply discard all candidates in $X_{k+1}$. Therefore,  $X=\bigcup_{i=1}^{k} X_i$ is also an $M_{\pi_X}$-XCC partition of $\pi_X$ where $M_{\pi_X} = M\cap X^2$. The induction hypothesis implies that $x_i >^\pi x_j$ for all $x_i \in X_i$ and $X_j\in X_j$ with $1 \leq i<j \leq k$. Combining with the fact that $x >^\pi y$ for all $x\in X$ and $y \in X_{k+1}$, we conclude that the theorem also holds for $k+1$.  
\end{proof}

\begin{theorem}
\label{t:unique-M-XCC}
For every anti-symmetric set $M\subseteq C^2$ of  constraints, the finest $M$-XCC partition exists and is unique. 
\end{theorem}
\begin{proof}
Let $M\subseteq C^2$ be an anti-symmetric set of constraints. The set of all $M$-XCC partitions is clearly nonempty since it contains the trivial partition $\{C\}$. The number of blocks of every $M$-XCC partition is at most $|C|$. Hence, there must exist an $M$-XCC partition with most blocks, that is, a finest $M$-XCC partition.  
For the uniqueness, consider two finest $M$-XCC partitions $\{X_1,\dots, X_k\}$ and $\{Y_1,\dots, Y_t\}$. We claim that $X_1=Y_1$. Indeed, if $X_1 \neq Y_1$ then either (1) $X_1\subsetneq Y_1$ or (2)  $Y_1\subsetneq X_1$ or (3) $X_1\setminus Y_1 \neq \varnothing$ and $Y_1\setminus X_1 \neq \varnothing$. 
\par 
\noindent 
\textbf{Case 1:} $X_1\subsetneq Y_1$. Since  both $\{X_1,\dots, X_k\}$ and $\{Y_1,\dots, Y_t\}$ are  $M$-XCC partitions, it is clear that $\{X_1, Y_1\setminus X_1, Y_2, Y_3, \dots, Y_t \}$ is also an $M$-XCC partition, contradicting the hypothesis that $\{Y_1,\dots, Y_t\}$ is a finest $M$-XCC partition. 
\par 
\noindent 
\textbf{Case 2:} $Y_1\subsetneq X_1$. We obtain a similar contradiction as in  Case 1. 
\par 
\noindent 
\textbf{Case 3:} $X_1\setminus Y_1 \neq \varnothing$ and $Y_1\setminus X_1 \neq \varnothing$. Let  $x\in X_1\setminus Y_1$ and $y \in Y_1\setminus X_1$. Thus $x\in Y_p$ for some $2\leq p \leq t$ and $y \in X_q$ for some $2\leq q \leq k$. 
In particular, $x \in X_1$, $y\in X_q$ with $q>1$. Since $\{X_1,\dots, X_k\}$ is an $M$-XCC partition, it follows that 
$$(x,y) \in M \text{ or else } \delta_{xy}>0 \text{ and } (y,x)\notin M \qquad (*)$$
Similarly, we have  $y \in Y_1$, $x\in Y_p$ with $p>1$. Consequently,  
$$(y,x) \in M \text{ or else } \delta_{yx}>0 \text{ and } (x,y)\notin M \qquad (**)$$
because $\{Y_1,\dots, Y_t\}$ is an $M$-XCC partition. 
\par 
\noindent 
\textbf{Case 3a:} $(x,y)\in M$. Then $(y,x)\notin M$ as and $M$ is anti-symmetric. Thus $(**)$ cannot be satisfied.  
\par 
\noindent 
\textbf{Case 3b:} $(y,x)\in M$. Similarly,  $(x,y)\notin M$ and $(*)$ cannot be satisfied.   
\par 
\noindent 
\textbf{Case 3c:} $(x,y), (y,x) \notin M$. Then $(*)$ implies that $\delta_{xy}>0$ while  $(**)$ implies that $\delta_{yx}>0$, a contradiction since $\delta_{xy}=-\delta_{yx}$. 
\par 
Note that the case $(x,y),(y,x)\in M$ does not occur since $M$ is assumed to be anti-symmetric. 
Therefore, we obtain a contradiction in all cases and the proof of the claim that $X_1=Y_1$ is complete. Let $M_{C\setminus X_1}=\{(x,y)\in M \colon x,y \notin X_1\}$. By the same argument applied to the  $M_{C\setminus X_1}$-Condorcet partitions $\{X_2, \dots, X_k \}$ and $\{Y_2, \dots, Y_t\}$ of the induced  election restricted to $C\setminus X_1$, we obtain $X_2=Y_2$. Thus by an immediate induction, we conclude that $X_1=Y_i$ for all $i$ and $k=t$. In other words, the two partitions $\{X_1, \dots, X_k \}$ and $\{Y_1, \dots, Y_t\}$ coincide and the proof is complete. 
\end{proof}

\section{Section 5.3: correctness and complexity of Algorithm 6 for the finest $M$-XCC partition} 
\label{s:algo-M-XCC}

\begin{lemma}
\label{t:index-XCC}
Let $M\subseteq C^2$ be an anti-symmetric and anti-reflexive set of  constraints. 
 For every $x\in C$, let $\gamma_x= |\Gamma_x|$ where 
 $$\Gamma_x = \{y \in C  \colon (x,y) \in M   \text{{ or else} } \delta_{xy}>0  \text{ and } (y,x)\notin M \}.$$ Let $\{X_1,\dots, X_k\}$ be an $M$-XCC partition. Let  $u\in X_i$ and $v\in X_j$ with $i<j$. Then $\gamma_u> \gamma_v$. In particular, if $\gamma_x=\gamma_y$ then $x,y\in X_i$ for some $i$. 
\end{lemma}

\begin{proof}
By the anti-symmetry of $M$,  it follows from the definition of $M$-XCC partitions that $\Gamma_u= (\Gamma_u \cap X_i) \cup \bigcup_{p=i+1}^k X_p$. 
Indeed, it is clear that $  (\Gamma_u \cap X_i) \cup \bigcup_{p=i+1}^k X_p \subseteq \Gamma_u$. For every $z\in X_l$ with $l<i$, we have $(z,u) \in M$ or else $\delta_{zu}>0$ and $(u,z)\notin M$. If $(z,u) \in M$ then by anti-symmetry, $(u,z) \notin M$ thus $z\notin \Gamma_u$. If $\delta_{zu}>0$ and $(u,z)\notin M$ then $\delta_{uz}=-\delta_{zu}<0$ and $(u,z)\notin M$ thus $z\notin \Gamma_u$. Hence, $\Gamma_u \cap X_l=\varnothing$ for all $l<i$. As $C= X_1\cup \dots \cup X_k$, we deduce that $  \Gamma_u \subseteq (\Gamma_u \cap X_i) \cup \bigcup_{p=i+1}^k X_p$. Consequently, $\Gamma_u= (\Gamma_u \cap X_i) \cup \bigcup_{p=i+1}^k X_p$. In particular, $ \gamma_u=|\Gamma_u| \geq \sum_{p=i+1}^k |X_p|$. 
\par 
Since $M\subseteq C^2$ is anti-reflexive and $\delta_{vv}=0$, we have $v\notin \Gamma_v$ and thus $|\Gamma_v\cap X_j|\leq |X_j|-1$. Note also that $\Gamma_v= (\Gamma_v \cap X_j) \cup \bigcup_{p=j+1}^k X_p$. Consequently, 
\begin{align*}
 \gamma_v =|\Gamma_v| & = |\Gamma_v\cap X_j| +   \sum_{p=j+1}^k |X_p| \leq |X_j|-1 + \sum_{p=j+1}^k |X_p| \\ & = -1+ \sum_{p=j}^k |X_p|< \sum_{p=j}^k |X_p| \leq \sum_{p=i+1}^k |X_p| \leq \gamma_u. 
\end{align*} 
where the second last inequality follows from  $j\geq i+1$. We conclude that $\gamma_v < \gamma_u$. The last statement is an immediate consequence.
\end{proof}

\begin{theorem}
\label{t:algo-finest-correct}
    For  every anti-symmetric and anti-reflexive input  set    $M\subseteq C^2$ of  constraints,  Algorithm~6 outputs the unique finest $M$-XCC partition.   
\end{theorem}

\begin{proof}
See Section 5.3 in the main paper. 
\end{proof}

\begin{theorem}
Algorithm~6 has time complexity $O(n^2)$ for every input constraints $M\subseteq C^2$ and every input NP matrix $[\delta_{xy}]_{x,y\in C}$ where $n=|C|$.  
\end{theorem}

\begin{proof} 
Note that computing $\Gamma_x$ and $ \gamma_x$ for each $x \in C$ requires $O(n)$ time; sorting $(\gamma_x)_{x\in C}$  to obtain $\textbf{X}^{0,M}$ requires $O(n\log n)$ time; computing $l$ and merging in the inner loop for the whole algorithm requires $O(n)$ time. Let $a_i= |X_i^{0,M}|$ then $a_1+\dots+a_w=n$. 
Suppose that we enter the outer loop while of Algorithm~6 with the current value $k$ and exit the loop with the next value $l+1$ of $k$.  Computing $V=\{ y \in C\setminus    \bigcup_{i=1}^k \widetilde{X}_i\colon \exists x \in \widetilde{X}_k, \, y \notin \Gamma_x\} $   
requires $O(a_k(a_{k+1}+\dots+a_w))$ time by checking the condition $y \notin \Gamma_x$ for each $y\in \bigcup_{i=k+1}^w \widetilde{X}_i$ and $x\in \widetilde{X}_k$.
 
Computing all the  updates of $V$ in the inner while loop requires \\  $O\left(\sum_{k+1 \leq i<l;\,  i <j \leq w } a_i a_j \right )$ time as we only need to check once the conditions  $y\notin \Gamma_x$  where $x\in \widetilde{X}_i$ and $y \in \widetilde{X}_j$ with  $k+1 \leq i<l$ and $i <j \leq w$.  
To summarize, each iteration of the outer while loop with entry value $k$ and exit value $l+1$ requires $O\left(\sum_{k \leq i<l;\,  i <j \leq w } a_i a_j \right )$ time. Summing over all  iterations, the time complexity of   Algorithm~6 is thus $O\left(\sum_{1\leq i<j \leq w} a_i a_j \right )$ which is in $O\left( \left(\sum_{1\leq i \leq w} a_i \right )^2\right)=O(n^2)$. The proof is thus complete. 
\end{proof}

\section{Section 7.1: estimation of the dispersion parameter $\theta_{MLE}$} 
 \begin{lemma}
     \label{l:theta-estimation} 
     With the  notations as in Section 7.1 in the main paper, the value   
     $
     \theta_{MLE}(V,\pi^*)
     $ is independent of the choice of the median   $\pi^*$ and is  denoted by $\theta_{MLE}(V)$. Moreover,  $\theta_{MLE}(V)$ is 
     is the unique solution  in $(0, 1]$ of the equation $H(\theta)=0$  where   
     $$
     H(\theta)= \sum_{k=1}^n \frac{k}{1-\theta^k} + \frac{n\theta}{1-\theta}-   \frac{n(n+1)}{2} - \frac{d_{V}^*}{m}. 
     $$
 \end{lemma}
 
  \begin{proof}
Recall that $
\theta_{MLE}(V,\pi^*) \in \mathop{\mathrm{arg\,max}}\limits_{\theta \in (0,1]}  P(X_1=\pi_1, \dots, X_m=\pi_m\vert \pi^*, \theta)$ 
where the random variables $X_1,\dots, X_m$ are  drawn independently from a Mallows distribution with  parameter $\theta$ and with the central ranking $\pi^*$.  
 By the independence of the variables $X_i$, the realization probability $P(V\vert \pi^*, \theta)$ of $V$ is given by 
 \begin{align*}
     P(V\vert \pi^*, \theta) & = P(X_1=\pi_1, \dots, X_n=\pi_n\vert \pi^*, \theta) \\& = \prod_{i=1}^m  P(X_i=\pi_i\vert \pi^*, \theta)   =\prod_{i=1}^m \frac{\theta^{d_{KT}(\pi_i, \pi^*)}}{Z_\theta} \\
     &  = \frac{\theta^{d_{KT}(\pi^*,V)}}{Z_{\theta}^m}= \frac{\theta^{d_{V}^*}}{Z_{\theta}^m} 
 \end{align*}
It follows that $\ln P(V\vert \pi^*, \theta) = d_{V}^* \ln \theta - m \ln Z_{\theta}$. Note that $\ln P(V\vert \pi^*, \theta)$ is maximized only if 
$\frac{d}{d\theta} \ln P(V\vert \pi^*, \theta)=0$, that is, 
\begin{equation}
\label{e:lnPV}
\frac{d_{V}^*}{\theta} - m \frac{d}{d\theta} \ln Z_{\theta} = 0
\end{equation}
We have the following formula (see, e.g., Lu and Boutilier (2014)): 
\begin{align} 
    \label{eq:constant-z} 
    Z_\theta & = (1+\theta)(1+\theta+\theta^2)\dots (1+\theta+\theta^2+\dots +\theta^{n-1}) \nonumber \\
    & = \prod_{k=1}^n \frac{1-\theta^k}{1-\theta}= \frac{1}{(1-\theta)^{n}}  \prod_{k=1}^n (1-\theta^k) 
\end{align} 
 \noindent 
It follows from \eqref{eq:constant-z} that $\ln Z_{\theta} = \sum_{k=1}^n \ln (1-\theta^k)- n \ln (1-\theta)$ and the condition \eqref{e:lnPV} becomes, after a  straightforward simplification, 
\begin{equation*}
      H(\theta)= \sum_{k=1}^n \frac{k}{1-\theta^k} + \frac{n\theta}{1-\theta} -   \frac{n(n+1)}{2} - \frac{d_{V}^*}{m}  =0, 
\end{equation*}
which is independent of $\pi^*$. The function $H$ is clearly  strictly increasing on $[0,1)$. The proof is thus complete. 
 \end{proof}

\end{document}